\documentclass{aa}
\usepackage{graphicx}
\usepackage{natbib}
 \usepackage{aalongtable}

\usepackage{txfonts}
\begin{document}
   \title{
   Chemical abundances for 11 bulge stars from high-resolution, near-IR spectra.
    \thanks{Based on observations collected at the European Southern Observatory, Chile 
(ESO Programme  079.B-0338(A))} 
}

\titlerunning{Chemical abundances of 11 Galactic bulge stars}
\authorrunning{Ryde, Gustafsson, Edvardsson, et al.}

   \author{N. Ryde
          \inst{1,2}
         \and
     B. Gustafsson
          \inst{2}
         \and
         B. Edvardsson
          \inst{2}
         \and
        J. Mel\'endez
          \inst{3}
         \and
       A. Alves-Brito
       \inst{4}
                \and
       M. Asplund
       \inst{5}
                \and
       B. Barbuy
       \inst{4}
                \and
       V. Hill
       \inst{6}
                \and
       H. U. K\"aufl
       \inst{7}
                \and
       D. Minniti
       \inst{8, 9}
                \and
       S. Ortolani
       \inst{10}
                \and
       A. Renzini
       \inst{11}
                \and
       M. Zoccali
       \inst{8}\\
          }

   \offprints{N. Ryde}

   \institute{Lund Observatory, Box 43, SE-221 00 Lund, Sweden
   \and
   Department of Physics and Astronomy, Uppsala University, Box 515, SE-751 20 Uppsala,
         Sweden
         \and
       Centro de Astrofisica da Universidade do Porto, Rua das Estrelas, 4150-762 Porto, Portugal
         \and
       Department of Astronomy, University of S\~ao Paulo, IAG, Rua do Mat\~ao 1226, S\~ao Paulo 05508-900, Brazil
         \and
Max-Planck-Institut f\"ur Astrophysik, Karl-Schwarzschild-Str. 1, D-85748 Garching
         \and
Boulevard de l'Observatoire, B.P. 4229, F-06304 NICE Cedex 4
         \and
ESO, Karl-Schwarzschild-Str. 2, 85748 Garching, Germany
              \and
 Department of Astronomy and Astrophysics, Universidad Catolica de Chile, Casilla 306, Santiago 22,
Chile
         \and
Vatican Observatory, V00120 Vatican City State, Italy         
        \and         
  Department of Astronomy, Padova University, Vicolo dell'Osservatorio 2, I-35122 Padova, Italy
 \and
     Osservatorio Astronomico di Padova, Vicolo dell'Osservatorio 5, I-35122 Padova, Italy\\
         \email{ryde@astro.lu.se}
             }

   \date{Received ; accepted }


 \abstract
{It is debated whether the Milky Way bulge has the characteristics of a classical bulge sooner than those of a pseudobulge. Detailed
abundance studies of bulge stars is a key to investigate the origin, history, and classification of the bulge. These studies can give constraints
on the star-formation
history, initial mass-function, and trace differences in stellar populations. Not many such studies have been made due to the large distance and large variable visual
extinction along the line-of-sight towards the bulge. Therefore, near-IR investigations are to be preferred.}
{The aim is to add to the discussion on the origin of the bulge and to study detailed abundances determined from near-IR spectra for
bulge giants already investigated with optical spectra, the latter also providing the stellar parameters which are very significant for the 
results of the present study. Especially, the important CNO elements
are better determined in the near-IR. Oxygen and other $\alpha$ elements are important for the investigation of the star-formation history. The C and
N abundances are important for determining the evolutionary stage of the giants but also the origin of C  in the bulge.}
{High-resolution, near-infrared spectra in the H band are recorded using the CRIRES spectrometer on the {\it Very Large Telescope}. The
CNO abundances can all be determined from the numerous molecular lines in the wavelength range observed. Abundances of the $\alpha$ elements Si, S,
and Ti are also determined from the near-IR spectra. } 
{[O/Fe],  [Si/Fe] and [S/Fe] are enhanced up to metallicities of at least [Fe/H]=$-0.3$, after which they decline. This suggests that the Milky Way
bulge experienced a rapid and early star-formation history like that of a classical bulge. However, a similarity between the bulge trend and the
trend of the local thick disk seems present. Such a similarity could suggest that the bulge has
a pseudobulge origin. The C and N
abundances suggest that our giants are first-ascent red-giants or clump stars, suggesting that the measured oxygen abundances are those the stars were born with.
Our [C/Fe] trend does not show any increase with [Fe/H] which could have been expected if W-R stars have contributed substantially to the C abundances. No "cosmic scatter" can be traced around our observed abundance trends; the scatter found is expected,
given the observational uncertainties.}
{}   
 \keywords{stars: abundances, Galaxy: bulge, infrared: stars }

\maketitle
%

\section{Introduction}

One great unsolved question in cosmology is how galaxies formed and acquired their stellar populations \citep[see for instance][]{renzini}. Two scenarios have been traditionally entertained for the formation of
bulges, with one  leading to ``classical" bulges, whereby
they form from merger-driven starbursts, and another leading to
``pseudobulges" that would result from the ``secular", dynamical
evolution of disks \citep{kormendy}. In the classical
bulge scenario most stars originate in a short phase of star formation
when the universe was only a few Gyr old, and bulge formation may
precede that of the disk, that could be acquired later. Instead, in the
so-called ``pseudobulges" stars form in the disk, over an extended
period of time, and the bulge results from the secular evolution of
the disk driven by the development of a bar. Thus, one expects
classical bulges to be made almost exclusively of old stars, while
stars in pseudobulges would have an age spread comparable to the
Hubble time.

The bulges of early-type spirals (Sa and Sb) are generally considered
to be ``classical", while pseudobulges are thought to inhabit
preferentially late-type spirals (Sc and Sd). The Sbc Milky Way (MW) galaxy
sits morphologically on the borderline, hence it is no surprise that
the origin of its bulge is currently a matter of debate.
What is especially puzzling with the MW bulge is that it looks
dynamically ``pseudo'' (with its boxy, peanut-shaped bar, Kormendy \&
Kennicutt 2004), while its stellar content is just what one expects for a
``classical'' bulge. Indeed, deep color-magnitude diagrams (CMD) of the
MW bulge show no detectable trace of stellar populations younger than
$\sim10$~Gyr (Ortolani et al. 1995; Zoccali et al. 2003)\nocite{ortolani,zoccali:2003}, a
conclusion for which compelling evidence has been recently provided by
the deep CMD obtained with HST for a proper-motion selected sample of
bulge stars \citep{clarkson}.

Such a  sharp distinction between these two scenarios of bulge formation
has been called into question recently: observations of disk
galaxies at redshift $\sim 2$ (lookback time $\sim 10$~Gyr) have
indeed revealed that such early disks have radically different physical
properties compared to local disks of similar mass. The early disks are
characterized by much higher velocity dispersion and gas fraction, and
harbor massive, highly star-forming clumps \citep{genzel:06,forster}. Thus, dynamical instabilities in such
early disks would occur with much shorter timescales (few $10^8$
years) compared to local disks, with a secular, but fast evolution of
such disks resulting in the early formation of a bulge \citep{genzel:08}.  In parallel to observations, theories have also envisaged
an early production of bulges via the migration and central
coalescence of gas-rich clumps in high-redshift disks \citep[e.g.,][]{immeli,elmegreen,carollo,bournaud}.

Determinations of detailed chemical compositions are key data for studies of the origin and
evolution of stellar populations, since
they carry characteristic signatures of the objects that enrich the interstellar
gas. Abundance ratios are sensitive
to the time-scales of star formation, to the initial mass function
(IMF), and may disclose relations
between different stellar groups, since different elements are
synthesized by different processes and
stars. For the bulge, a high ratio
of $\alpha$-element abundances relative to Fe is observed, suggesting that the
star-formation period was early and very
short \citep{lecureur,fulbright:07} and that the bulge formed more rapidly than the thin, and perhaps even the thick galactic disk  \citep{mcwilliam:08}. 

With stellar abundance ratios carrying such a {\it genetic}
information on the origin of different stellar populations, \citet{zoccali} 
compared the [O/Fe] ratios of bulge stars with those of
thin disk and thick disk stars from \citet{bensby:04}. Finding
bulge ratios systematically higher than disk ones, Zoccali et
al. argued against the bulge having been appreciably contaminated by
the migration of thick disk stars analogous to those in the solar
neighborhood.  A similar consideration was put forward by \citet{lecureur}, based on the much higher values of their [Na/Fe], [Mg/Fe]
and [Al/Fe] ratios in bulge stars compared to those in thin and thick
disk stars from \citet{bensby:05}, \citet{bensby:kol}, and \citet{reddy}.
However, differences in abundance ratios may be also due to
differences in systematic errors when the abundance analysis is done
by different groups, using different data and methods. For example, cool giants with crowded 
spectra are prone to many systematic uncertainties \citep[see, for instance,][]{santos:09} and comparing abundances derived from metal-rich, cool giants to those of solar-type dwarfs 
 could be uncertain due to relative systematics.

In this context, Mel\'endez et al. (2008) carried on an homogeneous
abundance analysis of bulge, thick disk and thin disk giant stars,
confirming the [O/Fe] trend found by Zoccali et al. (2006) for bulge
stars, but finding [O/Fe] ratios for thick disk stars much higher than
those derived by Bensby et al. (2004). Thus, in the study by Mel\'endez et al.
the high thick-disk [O/Fe]-ratios appear to be indistinguishable from those
of the bulge. This similarity of [O/Fe] ratios between bulge and
thick disk stars weakens the conclusion of Zoccali et al. (2006) about
the genetic difference between the bulge and the thick disk, although
it remains to be seen whether this similarity holds also for other 
[$\alpha$/Fe] ratios, that appear so different in Lecureur et
al. (2007).  If thick disk and bulge giants follow the same [O/Fe] vs [Fe/H] relation
the inference would be that both have formed "rapidly". The properties of
z$\sim2$ disks, with their high velocity dispersion and high star formation rate,
suggest that what we see there is thick-disk formation.

Determination of abundances for a large sample of red giant stars and planetary nebulae \citep[cf.][]{chiappini:09} in various bulge fields as well as in the inner galactic disk  will obviously provide a most powerful method to
constrain the chemical evolution and models of the bulge \citep{matteucci:99,silk}. 
A way to achieve this is by high-resolution, near-IR spectroscopy of red-giant
stars. In the IR, the obscuration
in the direction of the bulge is considerably reduced, offering us  the opportunity to go to heavily reddened regions. 
Furthermore,  near-IR spectra suffer
much less from line blending than spectra
at optical wavelengths, which makes it possible to safely trace the continuum and avoid abundance criteria marred with blending lines, so important in abundance
analysis. Moreover, only the IR offers, even within a small wavelength range,
all indicators necessary to
accurately determine the CNO abundances by
the simultaneous observations of many clean CO, CN and OH lines. 
Here, we present the first data from our VLT/CRIRES programme in which we
systematically study stellar abundance ratios
from different parts of the Galactic bulge.
In particular, we have aimed at the key elements C, N, and O, but also include some $\alpha$ elements.

\section{Observations}

\begin{table}
  \caption{Account of our observations.}
  \label{obslog}
  \begin{tabular}{l c c c c c  }
  \hline
  \noalign{\smallskip}
    Star$^a$ & R.A. (J2000) & Dec. (J2000) &  $H$ & $t^b_\mathrm{integration}$ & S/N \\
       & & & &  [s] & \\
  \noalign{\smallskip}
  \hline
  \noalign{\smallskip}
 \textrm {B3-b1} &  18 \,08 \,15.8 & -25 \,42 \,10 &11.3  & 2400 & 60 \\
  \noalign{\smallskip}
 \textrm {B3-b7} & 18 \,09 \, 16.9 & -25\, 49 \,28 & 11.6 & 4200 &  40\\
 \noalign{\smallskip}
 \textrm {B3-b8} & 18 \,08 \, 24.6 & -25\, 48 \,44 & 11.9 & 3840 &  90\\
 \noalign{\smallskip}
 \textrm {BW-b6} & 18 \,04 \,45.1 & -29 \,48 \,52 & 11.9 & 3840 & 60 \\
  \noalign{\smallskip}
 \textrm {BW-f6} & 18 \,04 \,56.1 & -29 \,48 \,59 & 12.0 & 4800 & 90 \\
  \noalign{\smallskip}
 \textrm {B6-f1} & 18 \,10 \,04.5 & -31 \,41 \,45 & 11.9 & 1920 &50 \\
  \noalign{\smallskip}
 \textrm {B6-b8} &18 \,09 \,55.9 & -31 \,45 \,46 & 11.9 & 3840 & 60\\
  \noalign{\smallskip}
 \textrm {B6-f7} & 18 \,03 \,52.3 & -31 \,46 \,42 & 11.9 & 1920 &50\\
  \noalign{\smallskip}
 \textrm {Arp\,4203} &  18 \,03\, 23.6 &  -30 \,01\, 59  & 9.2  & 200  & 80\\
  \noalign{\smallskip}
 \textrm {Arp\,4329} & 18 \,03\, 28.4  & -29 \,58 \,42 & 11.1 & 1800  & 95\\
  \noalign{\smallskip}
 \textrm {Arp\,1322} &  18 \,03 \,49.4 & -30 \,01 \,54   & 10.3 & 600   & 110 \\
 \noalign{\smallskip}
  \hline
  \end{tabular}
 \begin{list}{}{}
\item[$^{\mathrm{a}}$] The designation of the stars is adopted from \cite{lecureur}. 
\item[$^{\mathrm{b}}$] The total integration times are given by NDIT $\times$ DIT $\times$ NEXP $\times$ NABCYCLES $\times$ 2, see the CRIRES User's manual at http://www.eso.org/sci/facilities/paranal/instruments/crires/doc/
\end{list}
\end{table}

\begin{table*}
  \caption{Stellar parameters for the model atmospheres of our programme stars given as $T_\textrm{eff}/\log g/\mathrm{[Fe/H]}/\xi_\textrm{micro}$. }
  \label{par}
  \begin{tabular}{l c c c c c}
  \hline
  \noalign{\smallskip}
    Star &  \multicolumn{2}{c}{Stellar parameters$^a$ determined by}   &      \multicolumn{3}{c}{Stellar model parameters adopted in this paper}\\
    & \citet{lecureur} & M. Zoccali (2009), private comm.  &$T_\textrm{eff}/\log g/$ [Fe/H] $/\xi_\textrm{micro}$ &  [$\alpha$/Fe] & $\xi_\mathrm{macro}^b$ \\
    &  &   see also \citet{zoccali:08} &   & & FWHM in km\,s$^{-1}$ \\
     \noalign{\smallskip}
  \hline
 \noalign{\smallskip}
  \textrm {B3-b1} & $4300/1.7/-0.78/1.5$ & $4400/1.7/-0.60/1.3$ &  $4365/2.0/-0.73/1.5$ 
  & +0.3 & 5.0 \\
  \noalign{\smallskip}
  \textrm {B3-b7}   &  $4400/1.9/+0.20/1.3$  & $4350/1.6/+0.21/1.2$   & $4310/2.1/+0.06/1.6$ 
 & +0.08  & 4.8\\
  \noalign{\smallskip}
 \textrm {B3-b8}   &  $4400/1.8/-0.62/1.4$  & $4350/1.7/-0.65/1.4$   & $4250/1.5/-0.69/1.4$ 
 & +0.3  & 4.8\\
  \noalign{\smallskip}
 \textrm {BW-b6}   & $4200/1.7/-0.25/1.3$ &  $4450/1.9/-0.20/1.4$ &  $4340/2.2/-0.16/1.5$ 
 & +0.15 & 4.3 \\
 \noalign{\smallskip}
 \textrm {BW-f6}   & $4100/1.7/-0.21/1.5$ & $4400/1.8/-0.23/1.6$ &  $4150/1.5/-0.31/1.6$ 
 & +0.2 & 5.0  \\
 \noalign{\smallskip}
 \textrm {B6-f1} &   $4200/1.6/-0.01/1.5$ & $4250/1.9/-0.08/1.5$  & $4030/1.3/-0.08/1.5$ 
 & +0.1 & 5.3 \\
 \noalign{\smallskip}
 \textrm {B6-b8}    &  $4100/1.6/+0.03/1.3$ &  $4250/1.8/-0.06/1.4$  &  $3985/1.1/-0.14/1.3$ & $+$0.1  & 5.7 \\
 \noalign{\smallskip}
 \textrm {B6-f7}  & $4300/1.7/-0.42/1.6$  & $4450/1.8/-0.27/1.5$ &  $4315/1.9/-0.34/1.6$ 
 & $+0.2$ & 4.8\\
 \noalign{\smallskip}
  \hline
  \noalign{\smallskip}
    &  \multicolumn{2}{c}{Stellar parameters$^a$ determined by}   &      \multicolumn{3}{c}{Stellar model parameters adopted in this paper}\\
    &  \multicolumn{2}{c}{\citet{fulbright:07} \& used by \citet{ryde_bulb1}}  &$T_\textrm{eff}/\log g/$ [Fe/H] $/\xi_\textrm{micro}$ &  [$\alpha$/Fe] & $\xi_\mathrm{macro}^b$ \\
    &  & &   & & FWHM in km\,s$^{-1}$ \\
     \noalign{\smallskip}
  \hline
   \noalign{\smallskip}
  \textrm {Arp4203} & \multicolumn{2}{c}{$3902/0.5/-1.25/1.9$} &  $3815/0.35/-1.25/1.8$ & +0.35 
  & 6.2 \\
  \noalign{\smallskip}  
   \textrm {Arp4329} &\multicolumn{2}{c}{ $4197/1.3/-0.90/1.5$} &  $4153/1.15/-1.02/1.5$ & +0.35 
  & 5.8 \\
  \noalign{\smallskip}  
   \textrm {Arp1322} & \multicolumn{2}{c}{$4106/0.9/-0.23/1.6$} &  $4250/1.5/-0.16/1.5$ 
  & +0.15 & 6.2 \\
  \noalign{\smallskip}
  \textrm {Arcturus} &  \multicolumn{2}{c}{$4290/1.5/-0.5/1.7$} & $4280/1.7/-0.53/1.7$  & +0.3 & 3.7\\  
  \noalign{\smallskip}  
  \hline
  \end{tabular}
   \begin{list}{}{}
\item[$^{\mathrm{a}}$  The stellar parameters are given as  the effective temperature, $T_\textrm{eff}$, in K, the logarithmic surface gravity, $\log g$, in cgs units, and the microturbulence, $\xi_\textrm{micro}$, in km\,s$^{-1}$] 
\item[$^{\mathrm{b}}$  The macroturbulence is given as the FWHM used in the final convolution of the synthetic spectra fitted to include both effects of stellar macroturbulence and the instrumental profile.]
\end{list}
\end{table*}

In the {\it VLT} programme `Unveiling the secrets of the Galactic bulge: an infrared spectroscopic study
of bulge giants' we have as yet observed 8 bulge stars in the {\it H band}  with 
the CRIRES spectrometer \citep{crires,crires1,crires2}. CRIRES is a cryogenic echelle
spectrograph designed for  high spectral resolution,
near-infrared observations.  Adaptive Optics (MACAO - Multi-Applications Curvature Adaptive Optics) was used, enhancing the spatial resolution and the signal-to-noise ratio. The Adaptive Optics, which is only
feasible and available in the near-IR, has also the advantage to reject
diffuse starlight which may affect observations in regions of high star density.

The giants we have observed are chosen from the optical investigation of \citet{lecureur}, sampling three fields towards the Galactic bulge; at
($l$, $b$)= ($0^\mathrm o$,$-6^\mathrm o$), ($1^\mathrm o$,$-4^\mathrm o$) [Baade's Window], and
($5^\mathrm o$,$-3^\mathrm o$) [Globular Cluster NGC6553].  \citet{lecureur} analysed UVES/FLAMES spectra and derived abundances, but  the determinations for the  important C, N, and O
elements need improvement. The K giants are chosen half-way up along the red-giant branch (RGB), 
with 4000~K $\leq T_{\rm eff} \leq$ 4500~K, a
range where the molecular diagnostics are optimal and where adequate S/N
ratios can be achieved. Spectra of
these stars may be modelled more successfully than those of stars higher up
along the RGB, which makes the
abundance analysis much more reliable. The surface compositions are
characteristic of the gas from which the
stars once formed, with the exception of changes in C and N, the sums of which are, however, expected to be left unaltered by dredge-up of CN-processed material from the stellar interiors.   The $H$ magnitudes and the total integration times (ranging from 32 to 80 minutes) for each programme star are given in Table \ref{obslog}.   Our observations were performed 
during a period from May 2007 to October 2008. 

\citet{zoccali:08} determined an  iron-abundance distribution for the bulge from approximately 800 K giants (including the ones we have observed) in four fields toward the bulge with the VLT/FLAMES in the GIRAFFE mode at $R=20,000$. They find a clear gradient going to lower latitudes.  The iron distribution functions of the stars in the different fields range mainly from $-1.5<\mathrm{[Fe/H]}< +0.5$, but the peaks of the distributions lie at lower  [Fe/H] for  lower latitudes. All our eleven stars from three fields in the bulge have less than solar metallicity. 

The projected slit width on the sky was $0.30\pm0.01\arcsec$  yielding a spectral resolution of  $R \sim \lambda/\Delta \lambda = 70,000$ with 3.0 pixels per spectral resolution element.\footnote{The scale in dispersion direction is
0.10 \arcsec/pixel at the center of the order and $0.095\arcsec- 0.108\arcsec$/pixel over
the focal plane from  the long wavelength-side to the short. In the spatial direction the scale is 0.087 \arcsec/pixel.
This change of scale is due to the change of the beam-diameter induced
by the off-axis reflection at the Echelle grating which produces an
anamorphism. } This will allow us to 
resolve blends, define the continuum,
and adequately take care of telluric lines. 
In principle, with ideal adaptive optics and perfect image quality the
true effective defining entrance slit would be the diffraction limited
image of the star itself as delivered to the instrument.\footnote{In such
a case the spectrograph entrance slit would only be a technicality to
reduce the background and to establish geometric alignment, here mostly
compensating pointing errors.} CRIRES, however, does not reach this
limit due to the finite pixel size ($\sim1500\,$ms$^{-1}$ equivalent) and the limited
optical quality of the internal optics,  mostly due to the relatively
large ZnSe pre-dispersion prism. Indeed, the optical quality of the
complete pre-disperser is just marginally sufficient for the nominal
resolution of $R=100,000$. In addition, at the wavelength of interest here,
the adaptive optics does not deliver diffraction limited images, but only a core with
a halo. This implies that the finite slit width rules the effective spectral
resolution. During the commissioning of the CRIRES-MACAO system the effective
point-spread function was analyzed in great detail, and even in the
K-band under perfect conditions the enslitted energy fraction for the nominal
0.2" slit never exceeded 60\%, while with fainter stars and normal seeing
conditions this fraction was more like 40-50\%. For more details see
Table 1 in \citet{paufique}.

The wavelength range expected from the  `$36/-1/i$'-setting of CRIRES (i.e. in Echelle order 36) was $1539.3-1565.4$~nm over 
the detector arrays, consisting of a mosaic of four Aladdin III InSb arrays in the focal plane.
At order 36 the blaze function of the grating limits the throughput for
detector arrays \#1 and \#4 markedly. We, therefore,  concentrated our analysis on
detector arrays \#2 and \#3.
Data for the first and fourth detector arrays were used to check abundances when possible. This is, nevertheless, an improvement in wavelength coverage compared with the Phoenix spectrometer \citep{phoenix} at the Gemini telescope, an instrument that inspired the design of the CRIRES spectrometer, with a total wavelength range that corresponds to approximately one of CRIRES's detector arrays ($\Delta \lambda / \lambda = 0.5\%$). It should be noted that there are small gaps of approximately 2 nm between the spectra on the detector arrays. 

The approximate signal-to-noise
ratios (S/N)  per pixel of the observed spectra at 1554.8 nm (a well chosen continuum region in the third detector array) are also given in Table \ref{obslog}. The S/N per resolution element is close to a factor of 2 larger. The S/N is difficult to measure since the many lines make it difficult to find a large enough region of continuum from which to estimate it. The numbers given here are indicative. The S/N varies by a factor of two between the detector
arrays, mainly due to the blaze function. The S/N for the third detector array increases with wavelength  from 90 to 110\% of  the S/N at the reference wavelength. For the second detector  the  S/N varies linearly with wavelength  from approximately 70\% to 100\% of  the S/N at the reference wavelength of the standard setting, here chosen to be at 1557.3 nm. 

The observed data were reduced with the ESO standard pipe-line reduction package. 
The CRIRES pipeline is based on the general and coherent approach by ESO
using common routines, also employed in other instruments \citep{pipeline}.
The wavelength solution is based on a physical model approach for CRIRES \citep{wavesolution} using telluric
emission lines and the new infrared line catalog for ThAr hollow arc lamps \citep{thar}.
For the extraction of one-dimensional spectra the data taken for
different positions of the star along the slit (resulting from nodding
and dithering) are corrected in the usual way for glitches and bad
pixels and then rebinned in wavelength space before co-addition. The
final extraction is based on the "optimal extraction" method
which preserves the flux without sacrificing  S/N-ratio.
\footnote{Full details of the pipeline can be found under
http://www.eso.org/observing/dfo/quality/CRIRES/pipeline/pipe\_reduc.html}

In addition to these eight giants, we have reanalysed the three bulge giants presented in \citet{ryde_bulb1}.  These were also observed with CRIRES, during 
its  science verification, on 12 August 2006. The $H$ magnitudes and the total integration times are given in Table \ref{obslog}.  The observations and data are similar to those of our new eight  stars, although they were observed at $R=50,000$ and processed with routines in the reduction package
{\tt IRAF} \citep{IRAF}\footnote{IRAF is distributed by the National Optical Astronomy Observatory, which is operated by the Association of Universities for Research in Astronomy (AURA) under cooperative agreement with the National Science Foundation.}, in order  to retrieve one-dimensional, continuum normalised, and
wavelength calibrated stellar spectra. 


\section{Analysis}

We  analyze our spectra by modelling the stellar atmosphere and
calculating synthetic spectra for the observed spectral region. These are  thereafter convolved
 in order to fit the shapes and
widths of the lines, including the stellar macroturbulence and instrumental broadening.  We then derive elemental abundances by fitting the synthetic to the
observed spectra. In this section we will discuss the model
atmospheres, the stellar parameters including their uncertainties, and the spectrum synthesis.

\subsection{Model atmospheres}

We have calculated model atmospheres with the {\sc marcs} code \citep{marcs:08} after adopting the stellar effective temperature, logarithmic surface gravity, metallicity, microturbulence, and  [$\alpha$/Fe] enhancement for the model of each of our stars (see  Table \ref{par}, column 4).
The  {\sc marcs} standard  models are hydrostatic
and are computed on the assumptions of Local
Thermodynamic Equilibrium (LTE), chemical equilibrium, homogeneous spherically-symmetric stratification (in our case with $M_\mathrm{star}=0.8\,\mathrm{M_\odot}$),  and the conservation
of the total flux (radiative plus convective; the convective flux
being computed using the local mixing length recipe).
The  radiation field used in the model generation is
calculated with absorption from atoms and molecules by opacity
sampling at approximately 100\,000
wavelength points over the wavelength range $1300\,\mbox{\AA} $--$
20\,\mbox{$\mu$m}$. The models are calculated with 56 depth points from
a Rosseland optical depth of $\log \tau_\mathrm{Ross}=2.0$ out to $\log
\tau_\mathrm{Ross}=-5.0$. Data on absorption by atomic species are collected from the
VALD database \citep{vald} and Kurucz and other authors (for details, see Gustafsson et al., 2008\nocite{marcs:08}). The 
molecular-line opacity of CO, CN, CH, OH, NH, TiO, VO, ZrO,
H$_2$O, FeH, CaH, C$_2$, MgH, SiH, and SiO is included and
up-to-date dissociation energies and partition functions are used.
It should be noted that, for our targeted elements, we iterate and specify these abundances for each new iteration of the model atmosphere calculation, in order to be self-consistent.

The atomic line absorption files used in a  {\sc marcs}  model calculation are pre-calculated in a grid. Those used for our models are thus files with general metallicities closest to the metallicity
of the stars. The grid is given in steps of $\Delta \mathrm{[Fe/H]} = 0.25\,\mathrm{dex}$ in the relevant metallicity range.
For stars with [Fe/H]$\,<\,0.0$, [$\alpha$/Fe] can be solar or $\alpha-$enriched by up to $0.4\,\mathrm{dex}$.
For more metal-rich stars the abundance ratios are solar.
Microturbulence parameters of $1$ or $5\,\mathrm{km\,s^{-1}}$ were used in the model calculations.

  \begin{figure*}
   \centering
 \includegraphics[ width=\textwidth]{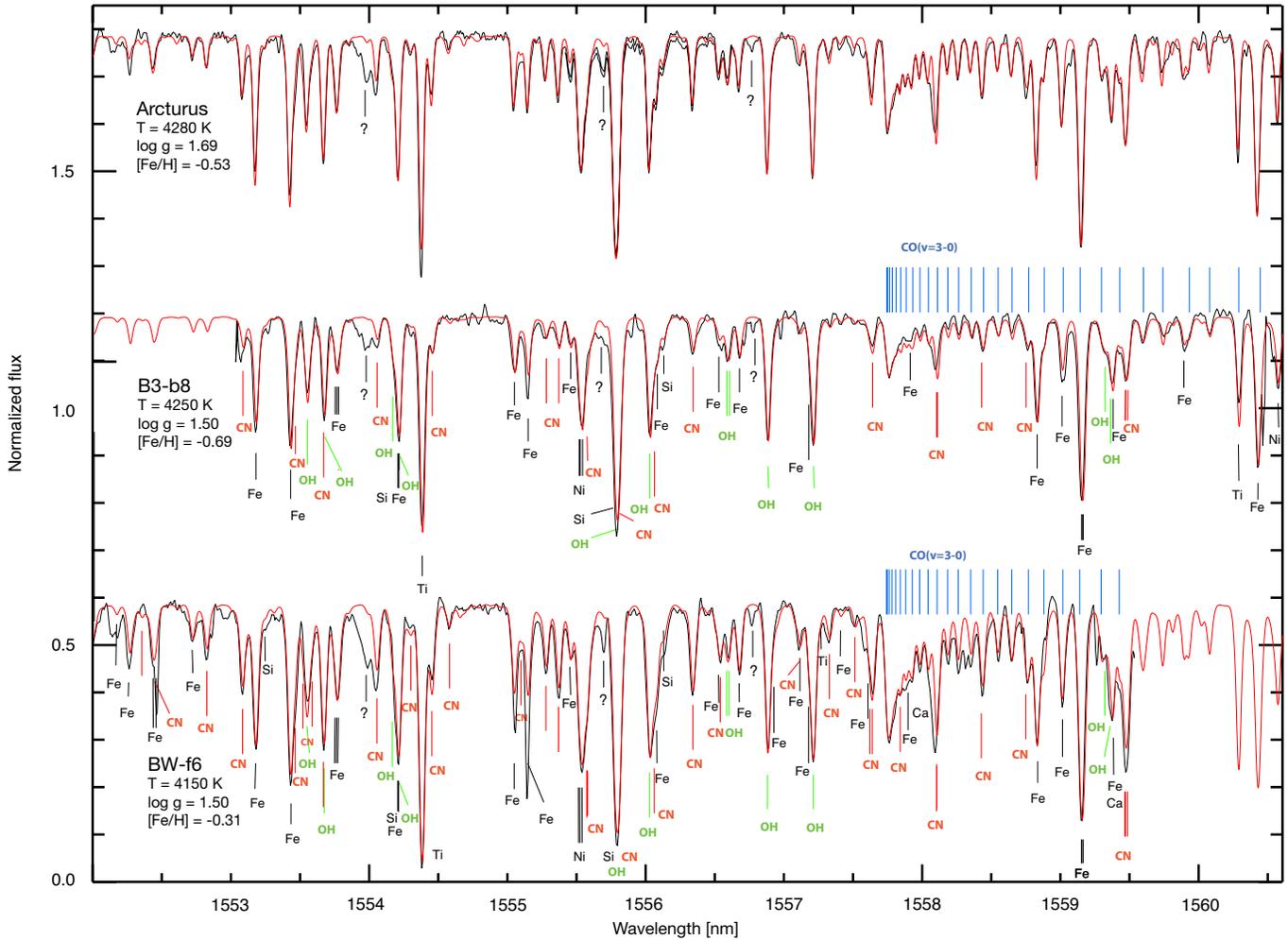}
      \caption{Sections of the observed CRIRES spectra of two of our  bulge giants  are shown with full, black lines. The observations are wavelength-shifted to laboratory wavelengths in order to enable a direct comparison between the different stellar spectra. Therefore, our observed spectra cover slightly different wavelength ranges. For comparison, the Arcturus atlas spectrum \citep{arcturusatlas_II} is also shown.
The parts of the spectra which have the highest signal-to-noise ratios are shown. From these parts  the C, N, and O elements can be determined. Our best synthetic spectra is shown in red. All synthetic
lines which are deeper than 0.97 of the continuum are identified. A few features are not identified in the Arcturus spectrum and are labeled with question marks. These features also show up in the bulge-star spectra.
A few lines in BW-f6 is conspicuously stronger than expected. They are probably affected by cosmic rays in the detector array.}
         \label{obs_model}
   \end{figure*}

\subsection{Fundamental stellar parameters and their uncertainties.}

\subsubsection{Deriving the fundamental parameters for our stars}

The fundamental stellar parameters, i.e $T_\mathrm{eff}$, $\log g$, [Fe/H], and $\xi_\mathrm{micro}$,  are needed as input for the model photosphere and spectrum synthesis.
Optical  \ion{Fe}{i} and \ion{Fe}{ii}  equivalent widths were used to
obtain spectroscopic surface gravities and effective temperatures by requiring a relative ionization and excitation
equilibrium with respect to a set of standard bright giants 
\citep[see ][]{melendez:2008}. The iron line list was carefully chosen to avoid significant blends
in K giants (Alves-Brito et al. 2009, in preparation), and is an extension of the line list already
presented in \citet{hekker}. This is
the same scale as used by \citet{melendez:2008}, which means that the stellar parameters for our
stars are  on the same scale as those of well-studied nearby giants, which ultimately
have effective temperatures determined from the Infrared Flux Method temperature scale of
\citet{ramirez2,ramirez1} and surface gravities based on absolute magnitudes as determined from Hipparcos parallaxes and adopting  stellar masses from isochrones.

The  equivalent widths ($W_\lambda$) of the iron lines for our new CRIRES stars
were taken from Lecureur et al. (2007), who obtained automatic $W_\lambda$ measurements 
from spectra observed with VLT/FLAMES in the UVES mode, providing a spectral resolution of $R=45,000$.  They used the DAOSPEC code which is described in \citet{stetson}. For one star, B6-b8, (which can be considered one of the most difficult cases, the star being cool and metal-rich) we also measured the equivalent widths manually. 

The microturbulence, $\xi_\mathrm{micro}$, was obtained
by requiring that the derived \ion{Fe}{i} abundances are independent of line strengths. This
procedure also yields the iron abundances. The [$\alpha$/Fe] used in our models are based on our preliminary
optical analysis of different $\alpha$-elements in bulge stars (Alves-Brito et al. 2009, in preparation), and so
it is based in a preliminary mean bulge relationship between [$\alpha$/Fe] and [Fe/H].

Our new stellar parameters for our bulge stars, including the three stars from \citet{ryde_bulb1},  and our reference star Arcturus, are given in 
Table \ref{par}. In the table we also 
give the [$\alpha$/Fe] values that we use in the determination of the parameters and later adopt in the model calculations. The  earlier determinations of the stellar parameters for these stars from the literature are also given in the table, namely those of  \citet{lecureur}, the updated parameters based on UVES data as described in \citet{zoccali:08}, 
and finally  \citet{fulbright:06}, the latter also used by \citet{ryde_bulb1}. The differences in the stellar parameters,  sometimes quite large, are mainly caused by the different methods applied in the different
stellar parameter determinations. For instance, the different line sets used 
and our relative ionization-equilibrium constraint to obtain the surface gravities, can cause differences.  The different values illustrate the general uncertainty, in particular the difficulties in determining  the fundamental parameters from the equivalent widths observed with UVES/FLAMES \citep{lecureur}.  Our new effective temperatures are systematically cooler (on average by $159$~K) than the updated parameters based on the UVES data, as described in \citet{zoccali:08}. For the three stars from \citet{fulbright:06} the differences go both ways with a maximum of $144$~K.

For the stars from \citet{ryde_bulb1}, which we analyse here too, we have redetermined the fundamental parameters based on the same scale as that used for other stars presented here.   We have also  redetermined the fundamental parameters for our reference star Arcturus ($\alpha$ Boo) in the same fashion. The new parameters for these stars are given in the lower part of Table \ref{par}. 
The changes in the derived C, N, and O abundances for these stars compared with those derived by  \citet{ryde_bulb1}  
agree with what is expected from the sensitivities of these abundances to the stellar parameters, and keeping in mind that different solar C, N, and O values are used in our paper compared to \citet{ryde_bulb1}.

It would be preferable to obtain accurate effective temperatures directly
from our IR spectra because of the substantial extinction towards the bulge
in the visual wavelength region.
This is, however, a difficult task. Although we are exploring the use of other infrared regions to improve the determination
of stellar parameters, the current infrared spectrographs available to us in 8m-class telescopes
only cover a narrow region, therefore requiring considerable amounts of telescope time to improve the
stellar parameters based solely on infrared data. 
Furthermore, the limited ranges in excitation energy for lines of various molecular
species limit their use for $T_{\rm eff}$ determination.
Effective-temperature sensitive features are particularly the OH
molecular lines which are used for oxygen abundance determinations. 
Other lines are only weakly sensitive.
Carbon, which appears in numerous lines of four different species, should
in principle be useful for $T_{\rm eff}$ determination.
Our tests show, however, that for a 100\,K increase in $T_{\rm eff}$
and for fundamental parameters which are typical of our targets 
the high-excitation C\,{\sc i} lines become only slightly stronger,
the CN lines stay unchanged, the CO lines become slightly weaker,
and the C$_2$ lines stay almost unchanged.
High S/N spectra or large wavelength regions would 
therefore have to be observed for the purpose. 
In the near future we, therefore, intend
to rely on spectroscopic equilibrium based on optical FeI and FeII lines. Besides the
current work of Alves-Brito et al. to define the zero-points of the spectroscopic stellar
parameters, we are currently acquiring more high resolution optical data of bright
K giants, which will allow us to further improve our stellar parameter scale. We are also
starting simulations that will allow us to estimate realistic uncertainties in the atmospheric parameters of cool giants
(Mel\'endez, Coelho, Alves-Brito, in preparation).

\subsubsection{Uncertainties in the fundamental parameters}

Our  $T_\mathrm{eff}$ values are uncertain, limited by the uncertainties in the $W_\lambda$ measurements. However, the small dispersion in the [O/Fe] vs. [Fe/H] plot (Figure \ref{ofe}) suggests that the random errors are of the order of  $75$\,K.   This is the value we use to derive the impact of this uncertainty on the derived abundances, see Table \ref{uncert}. However, the uncertainty could be underestimated.

Our spectroscopic surface gravity determinations are also very uncertain with uncertainties estimated to 0.3 dex (in some cases by as much as 0.5 dex), again mainly due to the quality of the $W_\lambda$ measurements of the few and weak Fe{\sc ii} lines but also the fact that we assume LTE in the ionization equilibrium, which might not be valid. However, our $\log g$ values
more or less follow our expectations for giants with $T_\mathrm{eff} = 4000$\,K ($\log g \sim1.0$) and
4300\,K ($\log g = 2.0$). On the other hand, these  gravities locate a few of the stars at distances not compatible with the bulge (all distances lie between 4.5 and 12 kpc). Our stars are chosen from the bulge giants of \citet{lecureur} which were selected to have a high probability of being bulge members. We have therefore also calculated the photometric surface gravities, using the PARAM tool \citep[see][]{dasilva}, forcing the stars to lie at a distance of 8\,kpc \citep{reid} and calculating the extinction in the same way as \citet{lecureur}. Our resulting gravities are given in Table \ref{grav} together with our spectroscopic ones. The $\log g$ values obtained by Zoccali et al. (and Lecureur et al.) are also determined photometrically and fall in the range $1.7\pm 0.1$~dex, in good agreement with our photometric values.  The main uncertainty in the photometric gravities is due to fact that 
the actual distances to the stars are not well known and the bulge does have an extension of a few kpc. Allowing a conservative uncertainty of $\pm2$\,kpc in the distance of $8$\,kpc, we find an uncertainty in the gravities of $0.25$\,dex, due to the distance uncertainty only.  The sensitivity of the determination of $\log g$ to the
extinction and the differential reddening, of which the latter is not taken into account, is small \citep[see][]{lecureur}. 
The gravities based on the two methods, given in Table \ref{grav}, are compatible with each other within the uncertainties, the differences being mostly $<0.3$\,dex, except for one case, B6-b8, which is a cool and metal-rich giant.  Fortunately, in our case, whichever $\log g$ determination we assume does not make a big difference, especially for the oxygen abundance from the OH lines, as can be seen from Table  \ref{uncert} and Figure \ref{ofe}. The carbon abundance determined from the CO lines is, however, affected more, see also Figure \ref{cnfe}. In the following discussion we have chosen to use the surface gravities determined spectroscopically.

\begin{table}
  \caption{Spectroscopically and photometrically determined surface gravities.}
  \label{grav}
  \begin{tabular}{l  c c c}
  \hline
  \noalign{\smallskip}
    Star  &   $\log g_\mathrm{spec}$   &    $\log g_\mathrm{photo}$  &  Difference  \\
    & & &  $\log g_\mathrm{photo}-\log g_\mathrm{spec}$\\
  \noalign{\smallskip}
    \hline
  \noalign{\smallskip}
  \textrm {B3-b1}   & 2.0 & 1.7 & $-0.3$ \\
  \noalign{\smallskip}
  \textrm {B3-b7}  & 2.1  & 1.8 & $-0.3$ \\
  \noalign{\smallskip}
 \textrm {B3-b8} & 1.5  & 1.7 & $0.2$ \\
  \noalign{\smallskip}
 \textrm {BW-b6}   & 2.2 & 1.9 & $-0.3$ \\
 \noalign{\smallskip}
 \textrm {BW-f6}    & 1.5 & 1.8  & 0.2 \\
 \noalign{\smallskip}
 \textrm {B6-f1}  & 1.3 & 1.6 & 0.3\\
 \noalign{\smallskip}
 \textrm {B6-b8} &   1.3  & 1.7 & 0.4\\
 \noalign{\smallskip}
 \textrm {B6-f7}   & 1.9 & 1.8 & $-$0.1\\
 \noalign{\smallskip}
  \textrm {Arp4203}   & 0.3   &  0.6 & 0.3\\
  \noalign{\smallskip}  
   \textrm {Arp4329} & 1.2  & 1.5 & 0.3 \\
  \noalign{\smallskip}  
   \textrm {Arp1322}   & 1.5 & 1.5 & $-$0.0\\
  \noalign{\smallskip}
 
  \hline
\end{tabular}
\end{table}


Finally, we estimate the uncertainty in the metallicity to be, in general, of the order of 0.05 dex and in the microturbulence to be of the order of 0.25\,kms$^{-1}$. 

The changes in the fundamental parameters due only to the two different measurements of the equivalent widths (automatic DAOSPEC or manually) 
of the cool and metal-rich giant,  B6-b8,
are 20 K in the temperature, 0.16 dex in $\log g$, $-0.17$ dex in  [Fe/H], and 0.35 km\,s$^{-1}$ in the microturbulence. 
The differences are within our estimated total uncertainties, except for the $\xi_\mathrm{micro}$ and the metallicity. This star is, however, particularly cool and metal-rich which means that the iron line measurements are especially difficult for this star.

\subsection{Synthetic spectra}

For the analysis of the observed spectra, we have generated
synthetic spectra, calculated in spherical symmetry for  our model photospheres. We sample the spectra with a resolution of 
$R=600,000$. With a microturbulence velocity of $1-2\,\mathrm{km\,s}^{-1}$, this will ensure an adequate sampling. In order to fit the observed spectra, we finally convolve our synthetic spectra with a macroturbulent broadening, represented by a radial-tangential 
function \citep{gray:1992}, and fitted to include both effects of macroturbulence and instrumental profile. The final macroturbulence parameters used are given in Table \ref{par}. The code used for calculating the synthetic spectra is BSYN v. 7.06 which is 
based on routines from the {\sc marcs}  code. Full consistency with the model atmosphere is achieved by choosing the same fundamental parameters, [$\alpha$/Fe],  individual abundances and in both calculations, including the molecular equilibria. A $^{12}\mathrm C/^{13}\mathrm C$ ratio of $ 24$ (96\% $^{12}\mathrm C$) is used for the bulge stars, and  of $9$ for Arcturus.

The atomic line-list used in our calculations is compiled from the VALD database \citep{vald} and from \citet{melendez:JH}.  The stellar parameters of our bulge stars resemble those of $\alpha$ Boo, which is therefore a good choice as a reference star. However, most of the lines used in our analysis can also be analyzed in the solar spectrum. We have, therefore, primarily checked the line list against the solar spectrum and corrected the line-strengths, if needed, by determining `astrophysical $\log gf$-values', fitting atomic lines in synthetic solar spectra to the observed one \citep{solar_IR_atlas}.  Hence, in our line list we have adjusted 96 lines based on the solar spectrum, see Table \ref{linelist}. The lines fitted were, among others, 
some Fe, Ni, Si, S and Ti lines. In addition, 4 Ti lines and 4 Si lines, which were too weak in the solar spectrum, were fitted to the $\alpha$ Boo spectrum from the  \citet{arcturusatlas}  atlas. These 8 lines are also given in  Table \ref{linelist}.  In order to determine the astrophysical  $\log gf$-values of these lines, we need to know the abundances of these elements in  $\alpha$ Boo.  Furthermore, apart from iron, Mg and Si are the most important electron donors at the continuum-forming regions, and therefore affect the line strengths through the continuous opacity (H$^-_\mathrm{ff}$). It is therefore important to estimate the abundance of Fe and Mg also, as well as possible.
Hence, the Fe, Mg, Si, and Ti abundances of our Arcturus modeling are based on the optically determined abundances derived by \citet{fulbright:07}, but taking into consideration our slightly different fundamental parameters, which, however, increased these abundances by only 0.00, 0.01,  0.03, and 0.01 dex, respectively. The other Fe, Si and Ti lines in our near-IR list, yield the same abundance for Arcturus, within a few tenth of a dex.
To conclude, the line list used here is similar to the one described in \citet{ryde_bulb1}, except that some new lines are added, a few omitted, the strengths of nine lines were adjusted slightly (less than 0.1 dex), and the strengths of 2 Ti lines ($15334.84$ and $15543.78$\,\AA ) and one Si line ($15506.98$\,\AA ) strengths  were adjusted by a larger amount. The latter three lines are all visible in the sun.


The molecular lists included are, for CO  \citet{goor},  SiO Langhoff \& Bauschlicher (1993)\nocite{lang}, CH \citet{jorg}, CN J\o rgensen \& Larsson (1990) and  Plez (1998, private communications)\nocite{jorg_CN}, OH \citet{gold}, and C$_2$ Querci et al. (1971) and J\o rgensen (2001, private communications)\nocite{querci}. For the molecules, the line lists were adopted as they are, leading to the following C, N, and O abundances for Arcturus from the spectra in the Arcturus atlas \citep{arcturusatlas}:
 $\log\varepsilon_\mathrm C =8.08\pm0.11$ (from CO lines), $\log\varepsilon_\mathrm N=7.64\pm0.09$ (from CN lines), and $\log\varepsilon_\mathrm O =8.70\pm0.13$ (from OH lines), which are in excellent agreement  with the values derived by \citet{ryde_bulb1} and by the optical work by \citet{Lecureur:phd}, who derived $\log\varepsilon(\mathrm{C})=7.96\pm0.10$, $\log\varepsilon(\mathrm{N})=7.74\pm0.10$, and $\log\varepsilon(\mathrm{O})=8.70\pm0.05$.

From our spectra, we have determined elemental  abundances from the CO [$v=3-0$] band, around 20 suitable CN lines, some 20 suitable OH [$v=4-2, 3-1, 2-0$] lines,  
and numerous Fe\,{\sc i} lines. Silicon can be measured from 2 or 3 lines, sulphur from 2 lines, and titanium only from 1 line, which makes the derived Ti abundance the most uncertain of the $\alpha$ elements.
We find the best fits, line by line, by synthesizing a grid of model spectra with incremental differences of 0.05 dex in the abundance sought for and finding the best fit by visual inspection. While for the OH and CN lines, and the atomic lines, every suitable line was inspected, for the CO band the entire band was fitted.  
In Figure \ref{obs_model} we present two examples of spectra of our 8 bulge stars, namely those of B3-b8 and BW-f6. These are spectra for which we obtained the highest S/N. 
Only the third detector array, providing the highest S/N, is
shown, but spectra from the other detector arrays were also used in the analysis. The reference spectrum of Arcturus is also shown in the Figure. 

\begin{table}
  \caption{Uncertainties in the derived abundances due to uncertainties in the stellar parameters.}
  \label{uncert}
  \begin{tabular}{l  c c c c c}
  \hline
  \noalign{\smallskip}
      &   $\Delta { \log\varepsilon(\mathrm C) }$   &  $\Delta { \log\varepsilon(\mathrm N) }$ &  $\Delta { \log\varepsilon(\mathrm O) }$&  $\Delta { \log\varepsilon(\mathrm S) }$    \\
  \noalign{\smallskip}
    \hline
  \noalign{\smallskip}
   $\Delta T_{\mathrm{eff}}=+75$~K   &  +0.04  & $+$0.06 &  +0.12 &  $-$0.05 \\
  \noalign{\smallskip}
   $\Delta \log g = +0.3$\,(cgs)  &  +0.10 & $-$0.03 &  +0.00 & +0.10 \\
  \noalign{\smallskip}
  $ \Delta \xi_\mathrm{micro} = +0.25\,\mathrm{km\,s}^{-1}$  &  $-$0.01  & $-$0.02 & $-$0.02 & $-$0.00 \\
  \noalign{\smallskip}
  $ \Delta \mathrm{[Fe/H]} = +0.05$ dex &  $+$0.02  & $+$0.04 & $+$0.04 & $+$0.00 \\
  \noalign{\smallskip}
  $ \Delta \mathrm{[\alpha/H]} = +0.1$ dex  &  +0.03 & +0.02 &   +0.04 &  +0.02\\
 \noalign{\smallskip}    
 
  \hline
\end{tabular}
\end{table}

\subsection{Uncertainties in the derived abundances}

The propagation of uncertainties in the stellar parameters to uncertainties in the C, N, and O abundances is presented in Table \ref{uncert}, based on the discussion in \citet{ryde_bulb1}.  These uncertainties are derived for a typical star of our sample, namely BW-f6. The uncertainty in the [$\alpha$/Fe] ratio is included but ambiguous since the different  $\alpha$-elements show different trends. We have adopted a general enhancement of [$\alpha$/Fe] $=+0.2$  in the model calculations and in the calculation of the synthetic spectra. 
We estimate the total internal uncertainties in the derived C, N, and O abundances 
to be approximately $\Delta A_\mathrm C = 0.11$,  $\Delta A_\mathrm N  = 0.09$,  and $\Delta A_\mathrm O = 0.13$ dex. As a comparison, the standard 
deviations in the determinations of the C, N, and O abundances from the many observed CO, CN, and OH lines for a given model are small, less than 0.05 dex. For example, for a given star and model atmosphere, the determination of the oxygen abundances  from each of the approximately 20 suitable OH lines provides a mean oxygen abundance with a standard deviation of 0.04 (line-to-line scatter) and a standard deviation of the mean of  0.01 dex. We note, however, that the
error in the CNO abundances generated by errors in the fundamental parameters will correlate, according to Table \ref{uncert}. E.g., 
an underestimated effective temperature will lead to underestimated abundances of C, N, as well as O. 
For the $\alpha$ elements, here represented by sulphur, we estimate $\Delta A_\mathrm S = 0.11$ dex, i.e of the same order as the molecular lines. We see, however, that the oxygen abundance suffers the largest uncertainty which is mainly due to the uncertainties in the effective temperature. 

The relatively low star-to-star scatter in the [O/Fe] vs. [Fe/H] plot
(see Figure \ref{ofe}) confirms that our error bar in the stellar parameters, especially $T_\mathrm{eff}$, are reasonable. Although our [O/Fe] values are sensitive to the $T_\mathrm{eff}$, they are not much affected by the uncertainties in $\log g$, compare Table \ref{uncert} and the two panels in Figure \ref{ofe}. On the other hand, the oxygen abundance for seven of our stars that were derived by \citet{zoccali} from the [O{\sc i}] line at 6300\AA\ is not much affected by the effective temperature but much more by the surface gravity. For instance, a change in $\log g$ of $+0.3$ dex yields a change in the oxygen abundances of $+0.13$ dex from the [O{\sc i}] line (cf. $\sim 0.00$ dex from the OH lines), whereas a change in the temperature of 100 K yields a change of $+0.02$ dex (cf. $+0.16$ dex from the OH lines). 

Other systematic uncertainties that could affect the abundance results include those due to, for instance, the continuum placement, the model atmosphere assumptions (such as the treatment of  convection and the assumption of spherical symmetry),  the uncertainties in the line strengths ($\log gf$ values), and the dissociation energies of the molecules. Furthermore, possible non-LTE effects  in the line formation of both atomic and  molecular lines could affect the result in a systematic way. In the future, only a full non-LTE  analysis of all  relevant atoms and molecules would be able to disclose the magnitude of these systematic uncertainties that our LTE analysis might be plagued by.
For abundance ratios, several uncertainties partly cancel leading to smaller uncertainties. The uncertainty due to the placement  of the continuum is estimated to be relatively small, less than 0.03 dex.

\begin{table*}
  \caption{[C/Fe]$^{\mathrm{a, b}}$, [N/Fe], [(C+N)/Fe],  [O/Fe], [Si/Fe], [S/Fe], [Ti/Fe], and [Fe/H] for our 11 bulge giants 
  and Arcturus. 
  }
  \label{abund2}
  \begin{tabular}{l  c c c c c c c c c }
  \hline\hline 
  \noalign{\smallskip}
    Star  &      [C/Fe]   & [N/Fe] & [(C+N)/Fe] & [O/Fe] &  [Si/Fe] & [S/Fe] & [Ti/Fe] & [Fe/H] \\
  \noalign{\smallskip}
  \hline
  \noalign{\smallskip}
 \textrm {B3-b1} &   0.090 & 0.030 & 0.079 & 0.39  & $-$ & 0.45 & 0.20 & $-$0.73 &  \\
 \noalign{\smallskip}
 \textrm {B3-b7}&    $-$0.11 & 0.43 & 0.065 &  $-$0.01 & 0.10 & 0.10 &0.10 & $+$0.16 &    \\
 \noalign{\smallskip}
 \textrm {B3-b8}&    $-$0.10 & 0.21 & $-$0.018 &  0.39 & 0.28 & 0.28 &0.13 & $-$0.67 &    \\
 \noalign{\smallskip}
 \textrm {BW-b6} &  0.13 & 0.27 & 0.16 & 0.43 & $-$ & $-$ & $-$ &$-$0.15 &  \\
\noalign{\smallskip}
   \textrm {BW-f6} &   0.050 & 0.24 & 0.095 & 0.28 & 0.29 & 0.24 & 0.19  & $-$0.30 &  \\
\noalign{\smallskip}
   \textrm {B6-f1} & $-$0.10 & 0.37 & 0.044 & 0.030 &  0.090 & 0.090 & $-$ &$-$0.070 &  \\
\noalign{\smallskip}
   \textrm {B6-b8} & 0.090 & 0.43 & 0.18 & 0.17 & 0.20 & $-$ & $-$ & $-$0.21    \\
\noalign{\smallskip}
   \textrm {B6-f7} &  0.14 & 0.36 & 0.19  & 0.57 & 0.33 &0.33 & 0.38 & $-$0.32  &   \\
\noalign{\smallskip}
   \textrm {Arp4203} & $-$0.66  & 1.03 & 0.37 & 0.040 & 0.25 & 0.30 & $-$0.050 & $-$1.25  \\
\noalign{\smallskip}
   \textrm {Arp4329} &  $-$0.11 & 0.28 & 0.00 & 0.41 & 0.45 &0.46 & 0.15 & $-$0.97\\
\noalign{\smallskip}
   \textrm {Arp1322} &  $-$0.090 & 0.38 & 0.054 & 0.24 & 0.20 &0.28 & 0.18 & $-$0.060\\
\noalign{\smallskip}
   \textrm {Arcturus} &  0.19 & 0.35 & 0.23 & 0.51 & 0.36 &0.26 & 0.17 & $-$0.53 \\
\noalign{\smallskip}
     \hline
  \end{tabular}
\begin{list}{}{}
\item[$^{\mathrm{a}}$]  $\mathrm{[X/Fe]} = \{\mathrm { \log\varepsilon(X)} - \mathrm { \log\varepsilon(Fe)\}_{star}} - \{\mathrm { \log\varepsilon(X) } - \mathrm { \log\varepsilon(Fe)\}_\odot }$.
\item[$^{\mathrm{b}}$] We have adopted the following solar abundances  (Mel\'endez et al. 2008): 
  $\mathrm { \log\varepsilon(C)=8.42}$, $\mathrm { \log\varepsilon(N)=7.82}$, $\mathrm { \log\varepsilon(O)=8.72}$, $\mathrm { \log\varepsilon(Fe)=7.50}$.
\end{list}
\end{table*}

\section{Results} 

\begin{figure}
   \centering
 \includegraphics[ angle=90,width=9cm]{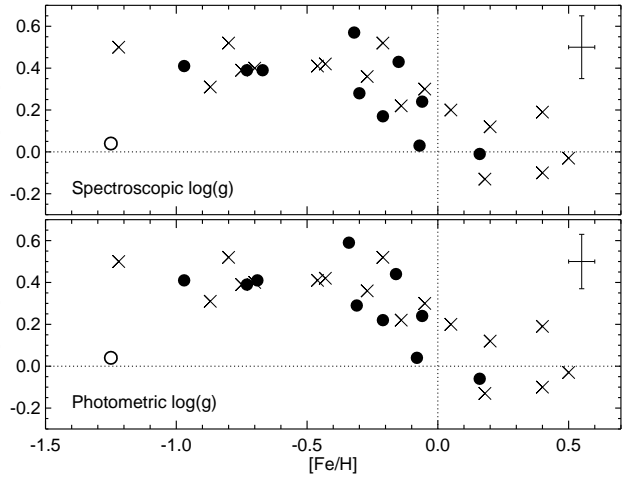}
      \caption{Logarithmic ratios of oxygen to iron normalized on the solar value for bulge stars.
Filled circles show our data from this paper, except the special giant  Arp4203 which is denoted by an open circle. 
Crosses show the results of Mel\'endez et al. (2008). Typical uncertainties are indicated in the upper right corner. The upper panel shows the oxygen abundances we derive when we use our spectroscopically derived $\log g$ values for our stellar atmosphere models, whereas the lower panel shows them when we instead use our photometrically derived surface gravities.}
         \label{ofe}
   \end{figure}

In Table \ref{abund2}  we present our derived  [C/Fe], [N/Fe], [O/Fe], [Si/Fe], [S/Fe], [Ti/Fe], and [Fe/H] for our eleven bulge giants. We also provide the derived C, N, C+N, O, and S abundances for Arcturus in addition to the Si, Ti and Fe abundances used.  Our  iron abundances of the bulge stars, as determined from 8 to 20 Fe-lines depending on the different S/N of the IR spectra, are systematically between 0.00 to 0.10 dex larger than the optically determined metallicity as given in Table \,\ref{par}, which is satisfactory.
In Figure \ref{ofe} we plot the [O/Fe] versus metallicity for our 11 bulge giants,  together with the bulge giants from \citet{melendez:2008}. These abundance ratios were also determined from near-IR spectra and the adopted stellar parameters were on the same scales. We find enhanced values of [O/Fe]$\sim+0.4$ up to approximately [Fe/H]$\sim-0.3$, after which they decrease. In the Figure we have especially marked the  giant Arp\,4203 which shows a large depletion of carbon, a large enhancement of nitrogen, and a [C+N/Fe] which is far from being solar, making this giant special. This was also noted by  \citet{fulbright:07}, \citet{melendez:2008}, and \citet{ryde_bulb1} who concluded that the oxygen abundance in this star should  probably not be used to represent the unprocessed [O/H] value for this bulge giant. 
 
\begin{figure}
   \centering
 \includegraphics[ angle=90,width=9cm]{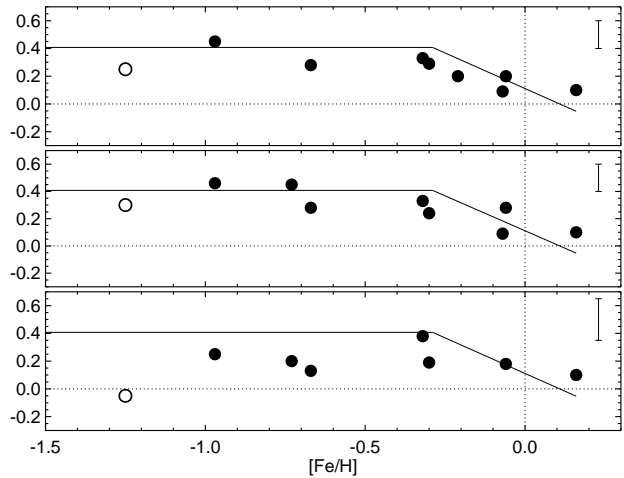}
      \caption{Logarithmic ratios of Si, S, and Ti to iron normalized on the solar value for our bulge stars are shown with filled circles. The [O/Fe] vs. [Fe/H] trend is indicated by a full line for reference in all panels. The measured values for the special giant Arp4203 are shown with an open circle. Typical uncertainties are indicated in the upper right corners.
}
         \label{afe}
   \end{figure}

In Figure \ref{afe} we plot our derived [Si/Fe], [S/Fe], and [Ti/Fe] versus our derived metallicity. For reference we also plot the trend we estimate from our [O/Fe] vs. [Fe/H] values. The abundances of the $\alpha$ elements are more uncertain than our derived C, N, O, and Fe abundances since there are much fewer lines to measure.  The uncertainties are therefore larger for these elements than for the C, N, O, and Fe abundances.  In spite of this,
we find that the [$\alpha$/Fe] values are enhanced for metallicities up to at least [Fe/H]$\sim-0.3$ after which they appear to decline. Below this metallicity [Si/Fe] and [S/Fe] are enhanced at a level of [$\alpha$/Fe]$\sim+0.3$. For higher metallicities they 
seemingly  decline and follow [O/Fe] for a given metallicity. [Ti/Fe] is generally lower for all metallicities.  Our $\alpha$ element trends overall corroborate the [O/Fe] enhancement trend, although for [Ti/Fe] at a lower value.  
In this Figure we have also especially marked Arp4203.

\begin{figure}
   \centering
 \includegraphics[ angle=90,width=9cm]{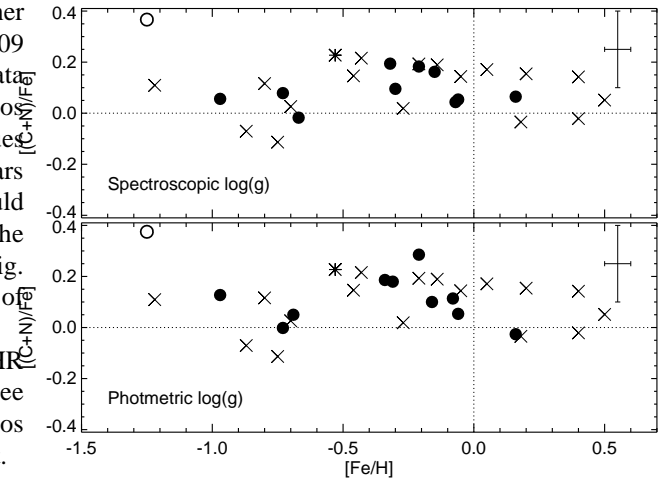}
      \caption{Logarithmic ratios of carbon+nitrogen to iron normalized on the solar value.
Our data are shown by filled circles, except Arp4203 which is shown by an open circle. Crosses show the results of Mel\'endez et al. (2008). The [(C+N)/Fe] value we derive for Arcturus is shown with a star. Typical uncertainties are indicated in the upper right corner.  The upper panel shows the values we retrieve when assuming spectroscopically derived $\log g$ and the values presented in the lower panel assumes photometrically derived surface gravities. }
         \label{cnfe}
   \end{figure}

All our stars except B3-b1 show significantly enhanced [N/Fe] values. If a star has experienced the first dredge-up, CN-cycled material is exposed at its surface. The abundances of C and N are then expected to change but their sum is left unaltered. In  Table \ref{abund2}   we also provide the calculated [(C+N)/Fe]. This is plotted in Figure \ref{cnfe} together with  the [(C+N)/Fe] for the bulge giants from \citet{melendez:2008}. These two data sets show approximately the
same pattern when it comes to the mean, standard deviation, and slope. We find a slope from a linear regression analysis for our data of $k=+0.07\pm0.09$ and for both sets of $k=+0.04\pm0.04$, i.e. both data sets are consistent with being sampled from a flat distribution. We further find a mean for both data sets of  $<$[(C+N)/Fe]$>=0.08\pm0.09$ (s.d.)  
and an error in the mean of $0.02$~dex.  Thus, both data sets show a systematic enhancement in the [(C+N)/Fe] ratios and they are therefore not consistent with being at solar values for all metallicities. Given our estimated uncertainties, our stars show no cosmic scatter. More stars and higher accuracy would be needed to judge
whether the off-set from solar values, the slight increase with
metallicity or the curved tendency in Fig. \ref{cnfe}  are real. We note that the highest value, apart from that of Arp4203, is that of Arcturus, with [(C+N)/Fe]=0.23.

\begin{figure}
   \centering
 \includegraphics[ angle=0,width=9cm]{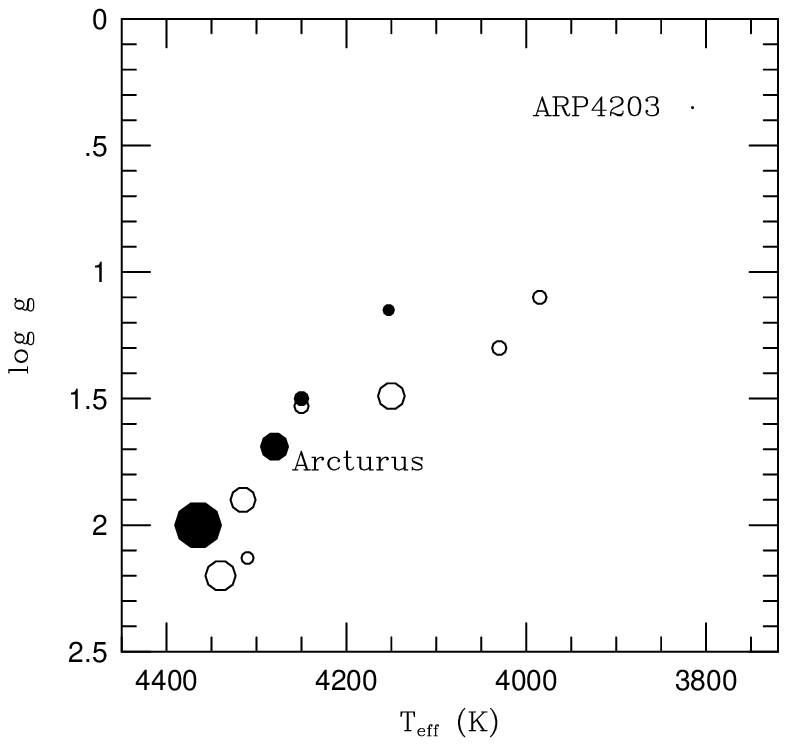}
      \caption{C/N ratios plotted in the theoretical HR
diagram. The dot diameters are proportional to C/N with a largest value of 4.6 
for B3-b1, 2.8 for Arcturus and 0.08 for Arp4203.
Stars with [Fe/H] $< -0.4$ are plotted as filled dots and the more
metal-rich ones as open circles.}
         \label{cn}
   \end{figure}

In Figure \ref{cn} we plot C/N for our stars  in the theoretical HR diagram  and   show how C/N varies with position in it. We see that the stars line up on the giant branch and that the C-N ratios decrease along it, as expected.  Arp4203 has evolved furthest.

\section{Discussion}

  \begin{figure}
   \centering
 \includegraphics[ angle=90,width=9cm]{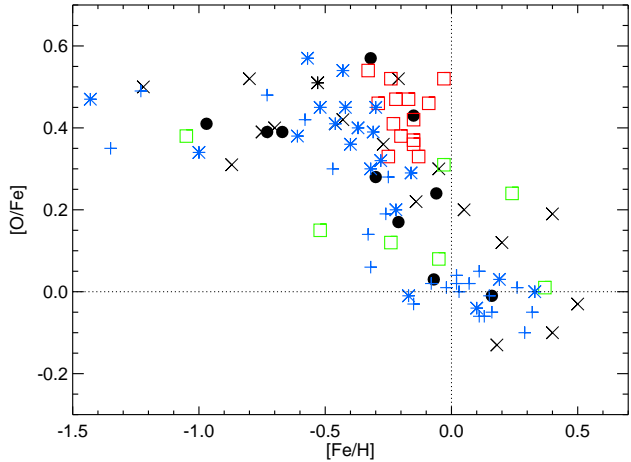}
      \caption{Logarithmic ratios of oxygen to iron normalized on the solar value. The oxygen abundances are here all determined from near-IR spectra.
Filled circles show our data from this paper, except that of Arp4203. Arcturus at [Fe/H]$ = -0.53$ is indicated with a black star at [O/Fe]=0.51, in agreement with that of \citet{Lecureur:phd}. Crosses show the results of Mel\'endez et al. (2008), green squares represent the abundances determined by \citet{cunha:2006}, and red squares  the data from \citet{origlia_M}. The blue star and plus signs represent thick disk and thin disk giants, respectively, from \citet{melendez:2008}. }
         \label{ofe_all}
   \end{figure}

\subsection{The [$\alpha$/Fe] trends}

The trends of the $\alpha$ elements (O, Mg, Si, S, Ca, and Ti) are of particular interest, since accurate 
[$\alpha$/Fe] ratios
in pre-AGB bulge stars will set strong constraints on the star formation
history.

\subsubsection{Oxygen}

Our oxygen abundances show a good $\it general$ agreement in the downward [O/Fe] vs. [Fe/H] trends with the results found by other authors. However, differences exist between these. In Figure \ref{ofe_all} we plot the [O/Fe] trends for the bulge stars based on analyses of near-IR spectra. These are, in addition to  our new results, values  from \citet{cunha:2006}, \citet{origlia_M}, \citet{melendez:2008}, and \cite{ryde_bulb1}. The [O/Fe] vs. [Fe/H] trend of \citet{origlia_M}  is moved upwards by 0.11 dex
to adjust to the authors' assumed solar oxygen abundance of 8.83 compared to our value of 8.72. 
It is a bit risky to quantify the similarities of the different trends due to the small number of stars, but assuming a constant [O/Fe] vs. [Fe/H] up to a metallicity of [Fe/H]$=-0.3$, and assuming a constant slope thereafter up to [Fe/H]$\sim+0.4$, we find an agreement in the slope of our data ($k=-1.0\pm0.3$) with  \citet{melendez:2008} ($k=-0.6\pm0.15$). Our data are also marginally consistent with the data of \citet{origlia_M}, which are confined to a narrow range in metallicity. The  \citet{cunha:2006} data set seems, however,  to suggest a more shallow slope ($k=-0.1\pm0.2$). 

\begin{figure}
   \centering
 \includegraphics[ angle=90,width=9cm]{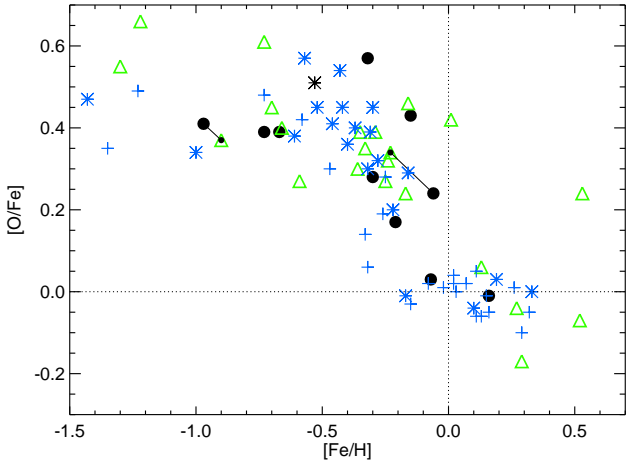}
      \caption{Logarithmic ratios of oxygen to iron normalized on the solar value. Filled circles show our data from this paper,  except that of Arp4203, and  a black star indicates Arcturus at [Fe/H]$ = -0.53$.
Triangles show the optically determined values by \citet{fulbright:07}. The two small dots represent the stars from  \citet{fulbright:07} which were analysed by \citet{ryde_bulb1} from near-IR lines and which are reanalysed by us. The determinations by Fulbright et al. and our determinations are connected by full lines. The blue stars and plus signs represent thick disk and thin disk giants, respectively, from \citet{melendez:2008}. }
         \label{ofe_all_optiskt_ful}
   \end{figure}

\begin{figure}
   \centering
 \includegraphics[ angle=90,width=9cm]{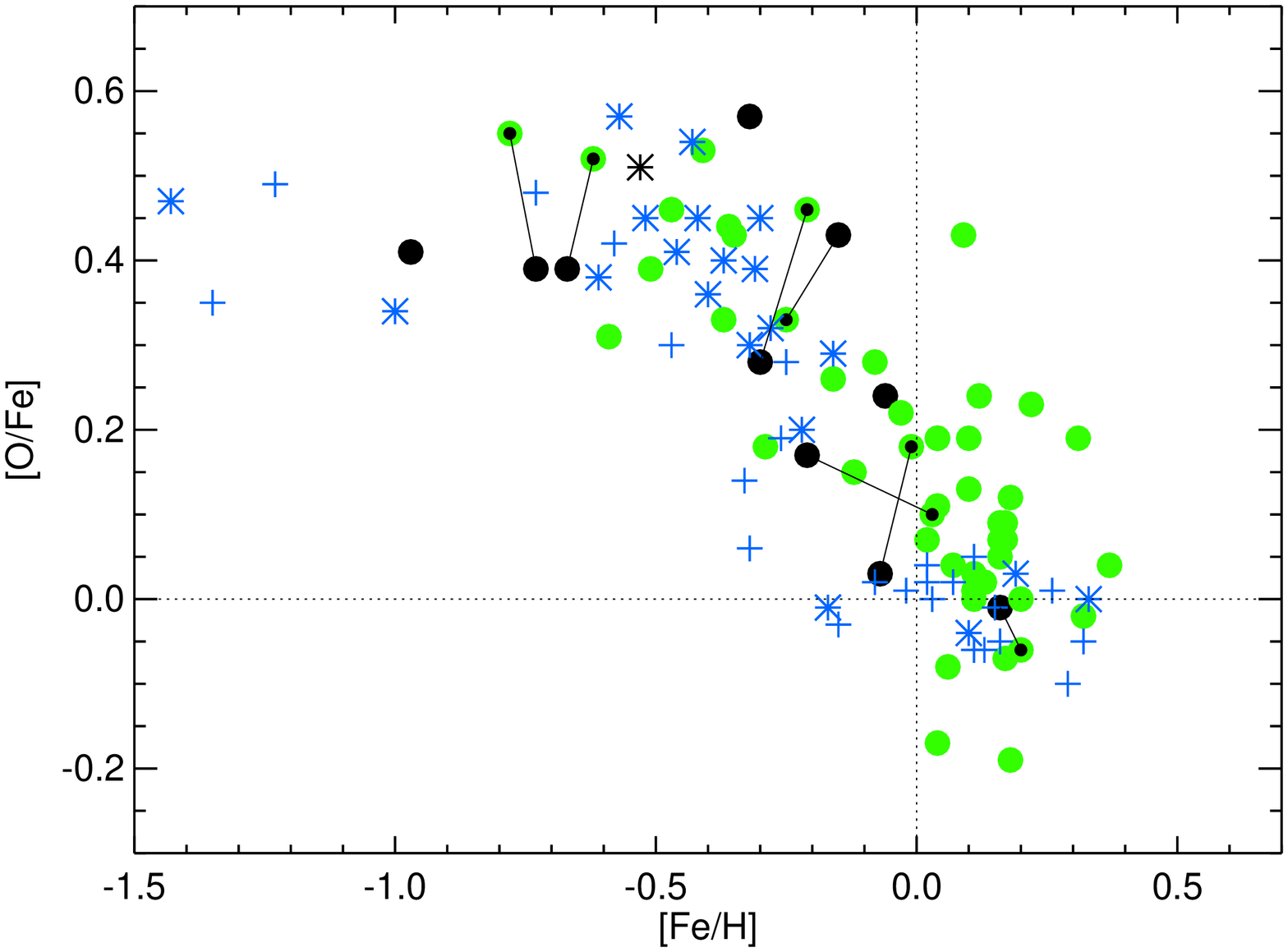}
      \caption{Logarithmic ratios of oxygen to iron normalized on the solar value. Filled black circles show our data from this paper,  except that of Arp4203, and  a black star indicates Arcturus at [Fe/H]$ = -0.53$. Filled green circles show the optically determined values by \citet{lecureur} and \citet{zoccali}. The small black dots mark the stars from  \citet{lecureur} and \citet{zoccali} for which we have determined the oxygen abundances from near-IR lines. These two determinations are connected. For the star B6-f7 there is only a near-IR determination of the oxygen abundance. When comparing these two determinations for the same stars, it should be noted that we have used new stellar parameters when determining the abundances from the near-IR spectra. The blue star and plus-sign symbols represent thick disk and thin disk giants, respectively, from \citet{melendez:2008}. }
         \label{ofe_all_optiskt}
   \end{figure}

In Figure \ref{ofe_all_optiskt_ful} we plot 
the data points from the optical work by \citet{fulbright:07}, which also show an agreement in the slope ($k=-0.8\pm0.15$. In Figure \ref{ofe_all_optiskt} we plot the optical results from \citet{zoccali}, also presented in \citet{lecureur} with a slope of $k=-0.6\pm0.15$, together with our determinations. Our results  are similar to these in scatter and slope. 

All our stars are in common with the optical analyses of \citet{zoccali} but for one of them it was not possible to determine an oxygen abundance  from the [O I] lines at 6300 \AA . 
The metallicities and [O/Fe] for these stars from our and their analyses are given in Table \ref{tab_zoccali}.  Note, that the stellar parameters are different in the two determinations.  
In Figure \ref{ofe_all_optiskt} we have marked and connected the two determination for the same stars.
When comparing the two analyses, we see that the metallicities are within the uncertainties, marginally also for B6-b8 ($\Delta \mathrm{[Fe/H]}_\mathrm{B6-b8} = 0.13$~dex). The oxygen abundances generally agree within uncertainties, with the largest difference being $\Delta \mathrm{[O/Fe]}_\mathrm{BW-f6} = 0.18$~dex. However, our abundances may tend to be systematically lower. Given that 
our OH lines are very temperature sensitive, one reason for the differences could be that the effective temperatures are still not determined accurately enough. Another reason could be  an overestimation of the line strengths in the optical spectra. Although the line strengths of the [O{\sc i}]  and the near-IR OH lines are comparable\footnote{For example, for the giant B3-b8 the [O{\sc i}] line has a strength of approximately $\log W_\lambda/\lambda=-5.1$ and the OH lines have line-strengths in the approximate range of $-5.7<\log W_\lambda/\lambda< -5.0$}, the optical spectra used  in  \citet{zoccali}  reaches S/N$\sim 50$ per resolution element, whereas our spectra have a S/N a factor of 2-4 larger and many more lines to use as oxygen criteria. 
Unknown blends 
might also affect the [O{\sc i}] line.

Our  [O/Fe] data suggest  a high value of +0.4 up to [Fe/H]$\sim-0.3$, after which the values decline rapidly with a slope of $k=-1.0\pm0.3$.  This is the mean trend we have plotted in Fig. \ref{afe}. A combination of  our [O/Fe] data and those of \citet{melendez:2008}  (see Figure \ref{ofe}) corroborates our finding. This trend also fits well with the values found by  \citet{origlia_M}  \citep[which are similar to those in][]{rich_07}  and consistent with the trends found by \citet{fulbright:07}.


In all the three comparison Figures we have also plotted [O/Fe] vs. [Fe/H] for the thin and thick disks from \citet{melendez:2008}.
When studying all  [O/Fe] determinations in Figures  \ref{ofe_all} to 
\ref{ofe_all_optiskt}, one gets the impression that these together may suggest similar, or possibly even higher values than those of the local thick disk of \citet{melendez:2008}. Note, that there may be important and different systematic errors in all these comparisons.


\begin{table}
  \caption{[Fe/H] and  [O/Fe] determined from near-IR lines in this paper and determinations based on optical lines \citep{lecureur,zoccali}, labelled L\&Z. 
  }
  \label{tab_zoccali}
  \begin{tabular}{l  c c c c c c}
  \hline\hline 
  \noalign{\smallskip}
    Star  &     [Fe/H] & [O/Fe]$^{\mathrm{a}}$ & [Fe/H]$_\mathrm{L\&Z}$ & [O/Fe]$_\mathrm{L\&Z}$\\
  \noalign{\smallskip}
  \hline
  \noalign{\smallskip}
 \textrm {B3-b1} &  $-$0.73 &   0.39  &    $-$0.78 &  0.55  \\
 \noalign{\smallskip}
\textrm {B3-b7}&      0.16 &  $-$0.01  &  0.20  & $-$0.06 \\
 \noalign{\smallskip}
 \textrm {B3-b8}&      $-$0.67 &  0.39 &  $-$0.62  & 0.52 \\
 \noalign{\smallskip}
 \textrm {BW-b6} &   $-$0.15 &  0.43 & $-$0.25 &  0.33\\
\noalign{\smallskip}
   \textrm {BW-f6} &     $-$0.30 & 0.28  &  $-$0.21 &  0.46\\
\noalign{\smallskip}
   \textrm {B6-f1} & $-$0.070 & 0.030 &   $-$0.01 &  0.18\\
\noalign{\smallskip}
   \textrm {B6-b8} & $-$0.10  & 0.10 &   $+$0.03 &  0.10  \\
\noalign{\smallskip}
   \textrm {B6-f7}  &  $-$0.32  &   0.57 & $-$0.42  & $-$\\
\noalign{\smallskip}
     \hline
  \end{tabular}
\end{table}

All trends seem to show a scatter that is similar or larger than the trends found for the thick and thin disks by \citet{melendez:2008}. This might, however,  be  expected for analyses of bulge stars since these are more difficult to analyse; our scatter reflects the expected uncertainties, and a cosmic scatter, if any,  must be smaller than that.

\subsubsection{Mg, Si, S, Ca, and Ti}

We find that [Si, S/Fe]$\sim+0.3$ for metallicities up to at least [Fe/H]$\sim-0.3$ above which they seem to decline.
Although Si and S may demonstrate slightly different trends, published abundance trends based on detailed abundance analyses for  bulge stars  all suggest that the ratio
of $\alpha$-element abundances relative to Fe show more or less enhanced values for all metallicities [Fe/H] $< 0.0$ \citep[e.g.][]{carretta,origlia_GC4,carnegie,origlia_GC3,origlia_GC2,origlia_GC1,origlia_M,cunha:2006,lecureur,fulbright:07,rich_07,mcwilliam:08,origlia:08}. 
For instance, \citet{lecureur} found high [$\alpha$/Fe] 
ratios in the bulge as compared to the disks, which  suggests an enrichment by mostly massive stars for all metallicities.  \citet{ryde_bulb1} measured sulphur (a product of explosive nucleosynthesis) from near-IR spectra and found enhanced values. \citet{carnegie} demonstrated that Mg and Si are enhanced by $0.3-0.5$ dex to super-solar metallicities. It should be noted here that our data do not support an enhanced [Si/Fe] and [S/Fe] values for [Fe/H$>-0.3$. The \citet{carnegie}  [Ti/Fe] and [Ca/Fe] trends show a steeper decline than for Mg and Si with metallicity but not as much as for oxygen. In order to fit the different trends they suggest a skewed initial mass-function (IMF) with more massive stars in the bulge. 
Later, \citet{fulbright:07} measured the abundances of O, Mg, Si, Ca, and Ti and found 
them all to decline in [$\alpha$/Fe] as a function of metallicity but also that they retain higher values than those of the disks for all metallicities. Mg was found to be enhanced the most, while Si, Ca, and Ti (which are thought to be products from the explosive nucleosynthesis phase of Type II supernovae) follow each other well at lower enhancement levels. The large decline of Si, Ca, and Ti compared to that of Mg was suggested to be caused by a metallicity-dependent decline of the former yields. 


Our results for the $\alpha$-elements do not support the existence of any significant "cosmic scatter" in the $\alpha$-element abundances
relative to iron in the bulge, but this is hardly conclusive since the observational scatter is considerable. 
 Alves-Brito et al. (2009 in prep.) show, however,
that their [$\alpha$/Fe] ratio in both their bulge and thick disk
giants have a scatter of only 0.03 dex. 
Also \citet{fulbright:07} find in their mean [$<$SiCaTi$>$/Fe] a small scatter, much smaller than that of halo stars, and  they interpret this as an indication that the bulge composition developed homogeneously, for example, due to efficient mixing.

\subsubsection{The thick disk-bulge similarity}

\citet{fulbright:07}, \citet{lecureur} and \citet{zoccali}  find abundance trends, including those for oxygen, that are different between the thick disk dwarfs and turn-off stars measured by Bensby et al. (2004) and Reddy et al. (2005) and bulge
 giants as measured from their own optical spectra. In contrast,
 \citet{melendez:2008}, by means of a homogeneous analysis of near-IR
 spectra of bulge and disk giants, find no chemical distinction between the local thick disk (up to [Fe/H]$\sim-0.2$) and the bulge, suggesting that the two populations show a similar chemical evolution and that the star-formation rates would not be significantly different. 
As shown in Figures  \ref{ofe_all} to \ref{ofe_all_optiskt}, the abundances measured here for bulge giants
 are consistent, within the uncertainties, with previous ones, both
 optical and near-IR. In the comparison with the thick disk, we follow
 the approach of Mel\'endez et al, restricting the comparison to giants
 in both components, measured in a fully consistent way. 
 Our bulge stars have abundances similar to those of thick disk giants.

\subsection{The carbon and nitrogen abundances}

Estimates from the optical wavelength region  of the carbon and nitrogen abundances are highly uncertain (in many cases only upper limits are known), whereas the CO, CN, and  OH lines  in the near-IR together easily provide them. The C and N abundances might give clues to, for instance,  the importance of W-R winds and the evolutionary state of the giants, thereby indicating whether the measured oxygen abundances are the original, unprocessed ones.

\subsubsection{First dredge-up and the measured oxygen abundance}

Low-mass giants that have ascended the giant branch for the first time
have only experienced the first dredge-up of CN-processed
material from the interior.  Thus, the CN cycle's products, which are mainly
 $^{14}\mathrm{N}$ and some $^{13}\mathrm{C}$ converted from  $^{12}\mathrm{C}$, are dredged-up to their
surfaces. This is expected not to alter the sum of the number of carbon and
nitrogen nuclei, while the measured oxygen
abundances should reflect the original abundances in the giants.
From Table  \ref{abund2}  we see that all stars (with the exception of Arp4203
) only show signs of the first dredge-up and thus no further
processing of oxygen through the NO cycle, nor any increase in the C abundance
characteristic of the third dredge-up on the Asymptotic Giant Branch.

The galactic chemical evolution of carbon and nitrogen are
still somewhat uncertain. \citet{bensby:kol}
find a constant [C/Fe] close to +0.1 for $-0.9<$[Fe/H]$<0.0$  for disk stars
and the summarized observational trend of [N/Fe] vs. [Fe/H]
as presented in \citet{prantzos:00} is constant at a
solar value. Thus, if the (C+N) abundances are expected to follow
that of iron, one would expect the [(C+N)/Fe] to be slightly below
+0.1 for all metallicities in the Galactic disk, which is consistent with what we find.
If the atmospheres of our stars had also been exposed to ON-cycled
material (in which $^{16}\mathrm{O}$ is converted to $^{14}\mathrm{N}$), their nitrogen abundances
would have been larger (resulting in larger [(C+N)/Fe]
ratios and a larger scatter) with an accompanying lower oxygen
abundance. It therefore seems likely that our measured oxygen
abundances can be taken as the starsÕ original unprocessed
abundance. This is also what is to be expected if the stars have
relatively low masses, and are still in the H-shell burning or He-core burning
phase, i.e. are on their first ascent along the giant branch or are clump stars.

\subsubsection{The role of W-R stars}

The oxygen abundance trends found from optical spectra of K giants in the bulge by
\citet{carnegie}, indicate a surprising interruption of oxygen production in the bulge
for high metallicities. 
The decrease in oxygen abundance is consistent with the 
strange scenario of no oxygen production for $\mathrm {[Fe/H]}> -0.5$.  \citet{carnegie} suggest that this could be connected to the onset of the Wolf-Rayet (W-R) phenomenon, which
would be vital for the production of the CNO elements.  Carbon is lost in  metallicity-sensitive, radiation-driven  stellar winds of  metal-rich W-Rs preventing carbon to be converted into oxygen, thereby reducing the oxygen production. Hence, the steep oxygen decline would  not be  specific for the stellar population(s) in the bulge, but a metal-dependent phenomenon with metallicity-sensitive stellar yields from massive stars 
playing an important role.  
Indeed, \citet{mcwilliam:08} show that the Galactic bulge and the thin disk experience the same decline in [O/Mg] versus [Mg/H] diagram, supporting this hypothesis  (or the alternative hypothesis that IMF is considerably metallicity dependent), since both oxygen and magnesium are synthesised in the hydrostatic cores of massive stars in a similar fashion.  In such a diagram the effects of the Fe-producing Type Ia supernovae are eliminated. The decline in the [O/Fe] vs. [Fe/H] plot in the bulge would also reflect the decrease in oxygen yields due to W-R stars, and not only the onset of Fe production from Type Ia supernovae. The latter gives clues to the timescales of the rates and duration of the star formation in the early bulge.   
Similarly, \citet{fulbright:07} also relate their low oxygen over-abundances to lower oxygen yields at higher metallicities, due to metallicity-dependent  W-R winds. This finds support in the calculations for rotating massive star models by
\citet{meynet:05} who conclude that higher C/O abundances are to be expected from high-metallicity WR stars. This idea is strengthened by  \citet{cunha:08} who derive Fluorine ($^{19}\mathrm{F}$) abundances for a sample of bulge stars. Their results suggest that winds from metal-rich W-R stars contribute more to the production of this element than do the AGB stars in the bulge compared to the situation in the disk.  

If the W-R hypothesis, invoked 
to explain the steeper decline of the [O/Fe] ratio  as compared to the other $\alpha$ elements is correct, it could mean a dramatic increase of the
carbon yields and thereby of the carbon abundances versus metallicity, since the carbon lost from the star is material that would
otherwise be expected to be transformed to oxygen in later stages and then expelled  by supernovae explosions. However, it is
fully conceivable that much of matter, lost in the W-R stage, which would else be transformed to oxygen, is primarily as yet helium. More
detailed model calculations are needed to explore which C enrichment is to be expected. The mass loss in massive stars would
anyhow have a large impact on the formation of carbon and oxygen, especially in metal-rich populations. The carbon
versus metallicity trend is therefore a crucial test of the W-R scenario. However, assuming that nitrogen is not affected we do not trace a dramatic increased carbon production from our data. The [(C+N)/Fe] ratio we find has a mean of $+0.08\pm0.09$ dex, and the  [(C+N)/Fe] vs. [Fe/H] shows a modest, if any, slope of $k= \frac{\partial\mathrm{[(C+N)/Fe]}}{\partial\mathrm{[Fe/H]}}=+0.07\pm0.09$. Thus,
our data do not give any strong further support to this hypothesis.







\section{Conclusions}

Abundance determinations for stars are known to be plagued by systematic errors that may be difficult to estimate.
In order to discuss the properties of different stellar populations, homogeneous differential spectroscopic studies, and detailed
comparisons of results of different studies, are significant. In the present study we have tried to follow this route, and find
a satisfactory agreement with results obtained in the optical, as well as IR, when a common temperature scale is used
for the stars. With our high-resolution IR spectroscopy, we have explored the CNO abundances, as well as the abundances of Si, S, Ti and Fe, for 11 bulge giants. We have found enhanced [O/Fe], [Si/Fe], and [S/Fe] values with increasing [Fe/H] up to approximately [Fe/H]$\sim-0.3$, after which these abundance ratios relative to Fe decrease. This suggests an early and rapid star formation in the bulge. Our investigation is not devised to make a detailed comparison with  thick disk stars and determine the relationship between these two populations; such a study should be made differentially to minimise systematic uncertainties.  Our abundance trends are, however, consistent with a similarity between these populations as was found in the differential study  by \citet{melendez:2008}. 
Such a similarity suggests that the picture of an isolated classical bulge may be oversimplified.
Inner-disk stars, at smaller galactocentric distances, should be explored in order to deepen the understanding of a possible physical connection between the bulge and the thick disk.

From our C and N abundances we conclude that our stars are first-ascent red-giants or clump stars, suggesting that their oxygen abundances are not affected by CNO cycling. Furthermore, we find that  there is no significant increase in the carbon abundances for high metallicities, which could have been expected if  W-R stars were to explain the large decline in [O/Fe] vs. metallicity.

We have demonstrated that  for the same stars several different determinations of the stellar parameters from optical spectra produce significantly different results, implying important systematic uncertainties. Attempts to reduce these should be made. Note also that  \citet{chiappini:09} compare, among others, the oxygen abundances derived from planetary nebulae (PNe) and giants in the bulge and find that the abundances determined from  giant star spectra are systematically higher by 0.3 dex. They conclude that this discrepancy may be caused by systematic uncertainties in either the PNe or giant star abundance determinations, or both.

To fully clarify the situation of the origin and evolution of the galactic bulge, additional near-IR abundance surveys of  elements (especially more $\alpha$ elements) are needed. Most earlier investigations have been restricted to Baade's window. Different regions of the
bulge are now being explored, and further systematic such work is needed.  

\begin{longtable}{llllcc}
\caption{\label{linelist} Line list of metal lines with astrophysical oscillator
strengths, see text.
(1) The wavelength in air,
(2) the excitation energy of the lower level,
(3) $\log gf$,
(4) the radiation damping parameter (when no value was available the very
small value of $1.00\times10^5$ was used),
(5) the van der Waals damping marked with an `A' when calculated according
to \citet{anstee}, \citet{barklem:00} and references therein,
or Barklem P., private communication (if a number is instead given this is
an empirical correction factor to the van der Waals damping computed
according to \citet{unsold}).
(6) a star `*' when the $\log gf$ value was determined from fits to the
spectrum of Arcturus rather than from the solar spectrum.}\\
\hline\hline
Wavelength [$\AA$] &  $\chi_\mathrm{exc}$ [eV]  & $\log gf$ & $\Gamma_\mathrm{rad}$ [rad\,s$^{-1}$] 
& van der Waal & Arcturus fit \\
\hline
\endfirsthead
\caption{continued.}\\
\hline\hline
Wavelength [$\AA$] & $\chi_\mathrm{exc}$ [eV] & $\log gf$ & $\Gamma_\mathrm{rad}$ [rad\,s$^{-1}$] & van der Waal & Arcturus fit \\
\hline
\endhead
\hline
\endfoot
 Si I  \\
 15330.191 & 6.718 & $-$1.90 & 1.00E+05 &    A   \\
 15338.780 & 6.261 & $-$2.58 & 1.00E+05 &    A   \\
 15342.973 & 7.108 & $-$1.85 & 1.00E+05 &    A   & * \\
 15361.160 & 5.954 & $-$2.12 & 1.00E+05 &    A   \\
 15375.430 & 6.734 & $-$1.53 & 1.00E+05 &    A   \\
 15376.830 & 6.223 & $-$0.66 & 1.00E+05 &    A   \\
 15376.830 & 6.721 & $-$1.13 & 1.00E+05 &    A   \\
 15381.738 & 6.721 & $-$2.03 & 1.00E+05 &    A   \\
 15387.069 & 7.166 & $-$1.64 & 1.00E+05 &    A   & * \\
 15471.964 & 6.726 & $-$2.40 & 1.00E+05 &    A   \\
 15496.964 & 7.006 & $-$2.54 & 1.00E+05 &  1.30  \\
 15506.980 & 6.727 & $-$1.55 & 1.00E+05 &    A   \\
 15520.115 & 7.108 & $-$1.85 & 1.00E+05 &    A   & * \\
 15532.449 & 6.718 & $-$2.18 & 1.00E+05 &    A   \\
 15538.463 & 6.761 & $-$2.36 & 1.00E+05 &    A   & * \\
 15557.790 & 5.964 & $-$0.65 & 1.00E+05 &    A   \\
 15561.251 & 7.040 & $-$1.23 & 1.00E+05 &    A   \\
 15638.472 & 6.734 & $-$1.93 & 1.00E+05 &    A   \\
 15674.653 & 7.064 & $-$1.30 & 1.00E+05 &    A   \\
 S I  \\
 15400.060 & 8.700 &  0.40 & 1.00E+05 &    A   \\
 15403.770 & 8.700 &  0.40 & 1.00E+05 &    A   \\
 15405.979 & 8.700 & $-$1.45 & 1.00E+05 &    A   \\
 15422.260 & 8.701 &  0.55 & 1.00E+05 &    A   \\
 15422.260 & 8.701 & $-$0.62 & 1.00E+05 &    A   \\
 15469.820 & 8.045 & $-$0.45 & 1.00E+05 &    A   \\
 15475.620 & 8.046 & $-$0.75 & 1.00E+05 &    A   \\
 15478.480 & 8.046 & $-$0.10 & 1.00E+05 &    A   \\
 Ti I  \\
 15334.840 & 1.887 & $-$1.05 & 2.93E+06 &    A   \\
 15381.110 & 2.333 & $-$2.29 & 7.13E+07 &    A   & * \\
 15399.285 & 2.334 & $-$2.00 & 7.13E+07 &    A   & * \\
 15426.970 & 1.873 & $-$2.52 & 2.76E+06 &    A   & * \\
 15543.780 & 1.879 & $-$1.15 & 2.76E+06 &    A   \\
 15602.840 & 2.267 & $-$1.60 & 2.31E+06 &    A   \\
 15698.979 & 1.887 & $-$2.14 & 2.76E+06 &    A   & * \\
 Cr I  \\
 15680.081 & 4.697 &  0.10 & 2.04E+08 &    A   \\
 Mn I  \\
 15673.385 & 5.133 & $-$0.57 & 6.68E+07 &    A   \\
 Fe I  \\
 15323.550 & 6.350 & $-$0.99 & 1.00E+05 &    A   \\
 15335.380 & 5.410 &  0.00 & 1.12E+08 &    A   \\
 15343.810 & 5.653 & $-$0.70 & 1.16E+08 &    A   \\
 15348.398 & 5.874 & $-$1.70 & 9.18E+07 &    A   \\
 15348.950 & 5.950 & $-$1.00 & 1.00E+05 &    A   \\
 15360.230 & 4.260 & $-$2.97 & 8.75E+06 &    A   \\
 15381.980 & 3.640 & $-$3.03 & 1.00E+05 &    A   \\
 15394.670 & 5.621 & $-$0.03 & 1.61E+08 &    A   \\
 15395.720 & 5.621 & $-$0.23 & 1.21E+08 &    A   \\
 15451.330 & 6.450 & $-$0.48 & 1.00E+05 &    A   \\
 15475.923 & 5.874 & $-$2.00 & 1.26E+08 &    A   \\
 15485.450 & 6.280 & $-$0.93 & 1.00E+05 &    A   \\
 15490.340 & 2.198 & $-$4.85 & 1.14E+04 &  1.40  \\
 15490.880 & 6.290 & $-$0.57 & 1.00E+05 &    A   \\
 15493.515 & 6.368 & $-$1.45 & 2.34E+08 &    A   \\
 15493.550 & 6.450 & $-$1.25 & 1.00E+05 &    A   \\
 15496.690 & 6.290 & $-$0.30 & 1.00E+05 &    A   \\
 15499.410 & 6.350 & $-$0.32 & 1.00E+05 &    A   \\
 15500.800 & 6.320 & $-$0.12 & 1.00E+05 &    A   \\
 15501.320 & 6.290 &  0.10 & 1.00E+05 &    A   \\
 15502.170 & 6.350 & $-$1.07 & 1.00E+05 &    A   \\
 15514.280 & 6.290 & $-$0.75 & 1.00E+05 &    A   \\
 15522.640 & 6.320 & $-$0.97 & 1.00E+05 &    A   \\
 15524.543 & 5.793 & $-$2.15 & 2.55E+08 &    A   \\
 15531.750 & 5.642 & $-$0.48 & 1.22E+08 &    A   \\
 15534.260 & 5.642 & $-$0.30 & 1.21E+08 &    A   \\
 15537.690 & 6.320 & $-$0.50 & 1.71E+08 &    A   \\
 15554.510 & 6.280 & $-$1.20 & 1.00E+05 &    A   \\
 15560.780 & 6.350 & $-$0.51 & 1.00E+05 &    A   \\
 15566.725 & 6.350 & $-$0.53 & 1.71E+08 &    A   \\
 15579.080 & 6.320 & $-$1.05 & 1.00E+05 &    A   \\
 15588.260 & 5.490 & $-$2.70 & 1.00E+05 &    A   \\
 15588.260 & 6.370 &  0.34 & 1.00E+05 &    A   \\
 15590.050 & 6.240 & $-$0.43 & 1.00E+05 &    A   \\
 15593.740 & 5.033 & $-$1.92 & 5.96E+08 &    A   \\
 15604.220 & 6.240 &  0.49 & 1.00E+05 &    A   \\
 15611.150 & 3.415 & $-$3.12 & 1.57E+07 &    A   \\
 15614.100 & 6.350 & $-$0.42 & 1.00E+05 &    A   \\
 15621.664 & 5.539 &  0.42 & 1.12E+08 &    A   \\
 15629.630 & 4.559 & $-$3.08 & 2.72E+08 &    A   \\
 15631.112 & 3.642 & $-$4.10 & 8.04E+07 &    A   \\
 15631.950 & 5.352 &  0.15 & 1.14E+08 &    A   \\
 15637.965 & 6.361 & $-$2.20 & 2.33E+08 &    A   \\
 15638.919 & 5.814 & $-$1.74 & 1.41E+08 &    A   \\
 15639.480 & 6.410 & $-$0.88 & 1.00E+05 &    A   \\
 15645.010 & 6.310 & $-$0.57 & 1.00E+05 &    A   \\
 15647.410 & 6.330 & $-$1.05 & 1.00E+05 &    A   \\
 15648.535 & 5.426 & $-$0.66 & 1.12E+08 &    A   \\
 15652.870 & 6.250 & $-$0.13 & 1.00E+05 &    A   \\
 15656.669 & 5.874 & $-$1.80 & 1.41E+08 &    A   \\
 15662.010 & 5.830 &  0.25 & 1.00E+05 &    A   \\
 15662.320 & 6.330 & $-$0.80 & 1.00E+05 &    A   \\
 15670.130 & 6.200 & $-$1.02 & 1.00E+05 &    A   \\
 15671.000 & 6.330 & $-$0.57 & 1.00E+05 &    A   \\
 15671.860 & 5.920 & $-$1.40 & 1.00E+05 &    A   \\
 15673.150 & 6.250 & $-$0.73 & 1.00E+05 &    A   \\
 15676.599 & 5.106 & $-$1.85 & 8.13E+07 &    A   \\
 15677.520 & 6.250 &  0.20 & 1.00E+05 &    A   \\
 15682.510 & 6.370 & $-$0.40 & 1.00E+05 &    A   \\
 15686.020 & 6.330 & $-$0.20 & 1.00E+05 &    A   \\
 15686.440 & 6.250 &  0.17 & 1.00E+05 &    A   \\
 15691.850 & 6.250 &  0.61 & 1.00E+05 &    A   \\
 15692.750 & 5.385 & $-$0.50 & 1.14E+08 &    A   \\
 Ni I  \\
 15555.370 & 5.488 &  0.13 & 1.48E+08 &    A   \\
 15556.016 & 5.283 & $-$3.15 & 1.68E+08 &    A   \\
 15605.680 & 5.300 & $-$0.47 & 1.00E+05 &    A   \\
 15605.680 & 5.300 & $-$0.96 & 1.00E+05 &    A   \\
 15632.654 & 5.305 & $-$0.01 & 9.71E+07 &    A   \\
\hline
\end{longtable}

\begin{acknowledgements}
N.R. is a Royal Swedish Academy of Sciences Research Fellow supported by a grant from the Knut and 
Alice Wallenberg Foundation. Funds from Kungl. Fysiografiska S\"allskapet i Lund are acknowledged.
NR, BE, and BG acknowledge support from  the Swedish Research Council, VR. JM acknowledge
support from the Portuguese FCT (PTDC/CTE-AST/65971/2006, Ciencia 2007). MZ and DM are supported by FONDAP Center for Astrophysics 15010003,
by BASAL CATA PFB 0609, and by FONDECYT. Kjell Eriksson is thanked for valuable help and discussions concerning the running of the MARCS programme. The referee is thanked for valuable suggestions.
\end{acknowledgements}

\bibliographystyle{aabib99}

\end{document}